\documentclass[fleqn,usenatbib,useAMS]{mnras}

\usepackage{graphicx}	% Including figure files
\usepackage{amsmath}	% Advanced maths commands
\usepackage{amssymb}	% Extra maths symbols
\usepackage{multicol}        % Multi-column entries in tables
\usepackage{bm}		% Bold maths symbols, including upright Greek
\usepackage{pdflscape}	% Landscape pages
\usepackage{rotating}
\usepackage[utf8]{inputenc} %Para tildes en autores

\newcommand{\kms}{\,km\,s$^{-1}$} % kilometres per second
\newcommand\Cp{{Ab1}}
\newcommand\Cs{{Ab2}}
\newcommand\Ct{{Aa}}

\newcommand\E{$\sigma$}

\usepackage[T1]{fontenc}
\usepackage{ae,aecompl}

\usepackage{newtxtext,newtxmath}
\title[Orbital properties of Herschel~36~A]{Spectroscopic study of the extremely young O-type triple system Herschel~36~A in the Hourglass Nebula. I. Orbital properties.}
\author[A. Campillay et al.]{Abdo R. Campillay$^{1}$\thanks{Contact e-mail: \href{mailto:acampillay@dfuls.cl}{acampillay@dfuls.cl}}, %\thanks{Present address: Universidad de La Serena, Av. Cisternas \mbox{1200}},
  Julia I. Arias$^{1}$,
  Rodolfo H. Barb\'a$^{1}$,
  Nidia I. Morrell$^{2}$,
  \newauthor
  Roberto C. Gamen$^{3},$
  Jesús Maíz Apellániz$^{4}$
\\
$^{1}$ Departamento de Física y Astronomía, Universidad de La Serena, Av. Juan Cisternas 1200 Norte, La Serena, Chile.\\
$^{2}$ Las Campanas Observatory, Carnegie Observatories, Casilla 601, La Serena, Chile.\\
$^{3}$ Instituto de Astrofísica de La Plata, CONICET--UNLP, and Facultad de Ciencias Astronómicas y Geofísicas, UNLP. Paseo del Bosque s/n, La Plata, Argentina.\\
$^{4}$ Centro de Astrobiología, CSIC-INTA, Campus ESAC Camino bajo del castillo s/n, 28 692 Villanueva de la Cañada, Spain.
}
  
\date{Last updated 2018 Sep 11; in original form 2018 Sep 11}

\pubyear{2018}

\begin{document}
\label{firstpage}
\pagerange{\pageref{firstpage}--\pageref{lastpage}}
\maketitle

% Abstract of the paper
\begin{abstract}
We present a detailed spectroscopic study of Herschel~36~A (H36A), the main stellar component of the massive multiple system Herschel~36 in the Hourglass Nebula, based on 
high-resolution optical spectra obtained along an 11 years span.
The three stellar components present in the spectrum of H36A are separated by means of a spectral disentangling technique. Individual spectral classifications are improved, and high precision orbital solutions for the inner and the outer orbits are calculated.
\noindent H36A is confirmed to be a hierarchical triple system composed of a close massive binary (Ab1+Ab2, O9.5\,V+B0.7\,V) in wide orbit around a third O-type star (Aa, O7.5\,Vz). The inner-pair orbit is characterized by a period of $1.54157\pm0.00006$ days, and semi-amplitudes of $181.2\pm0.7$ and $295.4\pm1.7$ km\,s$^{-1}$. The outer orbit has a period of $492.81\pm0.69$~days, and semi-amplitudes of $62.0\pm0.6$ and $42.4\pm0.8$~km\,s$^{-1}$. Inner and outer orbits are not coplanar, having a relative inclination of at least 20 degrees. 
Dynamical minimum masses of 
$20.6\pm0.8$~M$_{\sun}$, $18.7\pm1.1$~M$_{\sun}$, and $11.5\pm1.1$~M$_{\sun}$
are derived for the Aa, Ab1, and Ab2 components, respectively, in reasonable agreement with the theoretical calibrations. 
\end{abstract}
% Select between one and six entries from the list of approved keywords.
% Don't make up new ones.

%%%%%%%%%%%%%%%%%%%%%%%%%%%%%%%%%%%%%%%%%%%%%%%%%%

%%%%%%%%%%%%%%%%% BODY OF PAPER %%%%%%%%%%%%%%%%%%

% The MNRAS class isn't designed to include a table of contents, but for this document one is useful.
% I therefore have to do some kludging to make it work without masses of blank space.
%\begingroup
%\let\clearpage\relax
%\tableofcontents
%\endgroup

%########################################################### INTRODUCTION
\newpage
\section{Introduction}

Multiplicity is a %signature 
distinguishing characteristic of the massive stars. 
The results of the in-course large spectroscopic, imaging and interferometric surveys of massive stars, which have samples of tens to hundreds of objects, indicate that at least $70\%$ of the O-type stars in the Milky Way are multiple systems with, on average, two companions per central object \citep{sana17}.
The presence of a close companion strongly affects the stellar evolution through mass-exchange episodes and dynamical interactions. Thus, together with the mass, the mass loss rate, and the rotation, it determines the final destiny of the star. 
However, how binaries are formed is still not fully understood. 
In particular, the processes that lead to the formation of a massive short-period (few days) binary remain unknown.

Herschel~36  
($\alpha_{2000}=18^{\rm h}\,03^{\rm m}\,40^{\rm s}.321$; $\delta_{2000} = -24^{\circ}\,22'\,42''.86$; 
$V=10.3$, \citealt{sung00};
O7:\,V + sec, \citealt{sota14}) is the main source responsible for the ionization of the Hourglass Nebula, a blister-type \ion{H}{ii} region on the west side of Messier~8, one of the most studied Galactic star forming regions. 
A distance of $1.25$~kpc for Herschel~36 (H36) was estimated by \citet{arias06}.
The recent parallax measurement of H36 ($0.90\pm0.22$ mas) published in the Data Release 2 of the Gaia mission \citep[GDR2]{gaia18a} also places the system at a similar distance from the Sun.

As shown by HST/WFPC2 images (\citealt{maiz15a}),
H36 can be considered a Trapezium-like system with no less than ten stellar components in a radius of $\sim 4''$, most of them discovered through the various infrared (IR) studies of the region (e.g. \citealt{woodward86,woodward90,stecklum95,arias06,goto06}).
Observational evidence, such as the compact morphology of the ionized gas emission and the high fraction of IR-excess sources, indicates the Hourglass region is extremely young. 
Herbig-Haro objects and other signatures of active star formation have also been found in this area (\citealt{arias06,arias07}).

The optically brightest component of the "H36 Trapezium" is H36A, the main target of this paper.
Among the most conspicuous secondary components, we mention the massive binary KS1 (H36B; 
\citealt{allen86,woodward90}),
the extended IR source Herschel 36 SE (H36C; \citealt{goto06}),
and the ultracompact HII region G5.97-1-17 (H36D), whose proplyd nature was established by \cite{stecklum98}.
All of these peculiar objects lie within $\sim3''$ from H36A, equivalent to 3750 au at the Hourglass distance.

Previously considered a single O star, H36A was demonstrated to be a spectroscopic triple system (SB3) by \cite{arias10}.
A picture of a hierarchical system, composed of a close massive binary (Ab1+Ab2; $1.5$~days period) in wide orbit ($\sim$500~days period) around a third O-type star (Aa) was proposed by these authors, who also determined preliminary radial velocity solutions for the inner and the outer orbits. 
After that, two components, namely the star Aa and the binary Ab, were resolved using the near-IR instrument AMBER attached to the VLT Interferometer \citep{sanchez14}.
In the latter study, the authors determined a projected distance of 1.8 mas, and a flux ratio between components close to unity.

This paper continues and supersedes the work from \cite{arias10} and is devoted to characterize the orbital properties of the triple system H36A in the most accurate possible way. 
The observational dataset is presented in Sec.~\ref{observations}. 
In Sec.~\ref{results} we describe the separation of the composite spectrum, the spectral classification of the individual stellar components, and the determination of the orbital solutions. 
These results are discussed in Sec~\ref{discussion}. 
Finally, Sec~\ref{outlook} summarizes the conclusions.

%########################################################### OBSERVATIONS
\section{Observations}
\label{observations}

The observations used in this work were obtained in the framework of the {\em OWN Survey}
(\citealt{barba10,barba17}), a high-resolution spectroscopic monitoring of massive stars. 
The {\em OWN Survey} started in 2005 and one of its goals is to set the multiplicity status of the Galactic O- and WN-type stars in the Southern hemisphere through radial-velocity studies. 
The survey makes use of astronomical facilities in Chile and Argentina, focusing on the \'echelle spectrographs attached to 2 m-class telescopes. 
For the particular case of H36A, the spectra were obtained at Las Campanas Observatory (LCO) and ESO/La Silla Observatory, both located in northern Chile, between 2005 to 2016.
Table~\ref{tdata} presents a summary of the data, including instrumental configurations, representative resolving powers ($R$), and typical signal-to-noise ratios (S/N). The last column of the Table quotes the number of spectra ($n$) acquired with each configuration.
Spectra from LCO were extracted and normalized using the standard IRAF \footnote{IRAF is distributed by the National Optical Astronomy Observatories, which are operated by the Association of Universities for Research in Astronomy, Inc., under cooperative agreement with the National Science Foundation.} routines, whereas FEROS spectra from La Silla Observatory were reduced by means of the standard pipeline in MIDAS package provided by ESO.

\begin{table}
  \begin{center}
  \caption{Observational data of Herschel 36 A.}
  \label{tdata}
  \begin{tabular}{lcccccc}
  \hline
  Observatory & Telescope & Spectr. & $R$ & S/N & $n$ \\
  \hline
  LCO      &  2.5 m "du Pont" & Échelle& 25000& 100 & 65 \\
  %LCO      &  6.5 m "Clay"    & MIKE   & 60000& 200 & 2  \\
  La Silla &  2.2 m MPI       & FEROS  & 48000& 150 & 24 \\
  \hline
  \end{tabular}
  \end{center}
\end{table}

%########################################################### ANALYSIS
\section{Analysis and Results}
\label{results}

\subsection{Spectral disentangling and radial velocities}
\label{separation}

To separate the composite spectrum of H36A we used the disentangling method by \cite{gl06}, an iterative procedure that allows to reconstruct the individual spectra of the stellar components of spectroscopic binaries and to compute their radial velocities. 
We applied a modified version of the code developed by \cite{veramendi12} for the treatment of triple and higher order spectroscopic systems.

For a thorough description of the technique the reader is referred to \cite{gl06} and \cite{veramendi12}, but the basic idea is to use alternately the spectrum of each component to calculate the spectra of the others. 
The single-lined spectra resulting from each step are then used to compute the radial velocities (RVs). 
The procedure is carried out in two blocks. 
The first one obtains the spectra of each component using the standard {\sc iraf} tasks for spectra manipulation, and the second one measures the RVs by cross-correlation. 
Each block depends on the results of the previous one. 
The procedure is thus iteratively run until convergence.

\begin{figure*}
\centering
\includegraphics[width=0.5\textwidth]{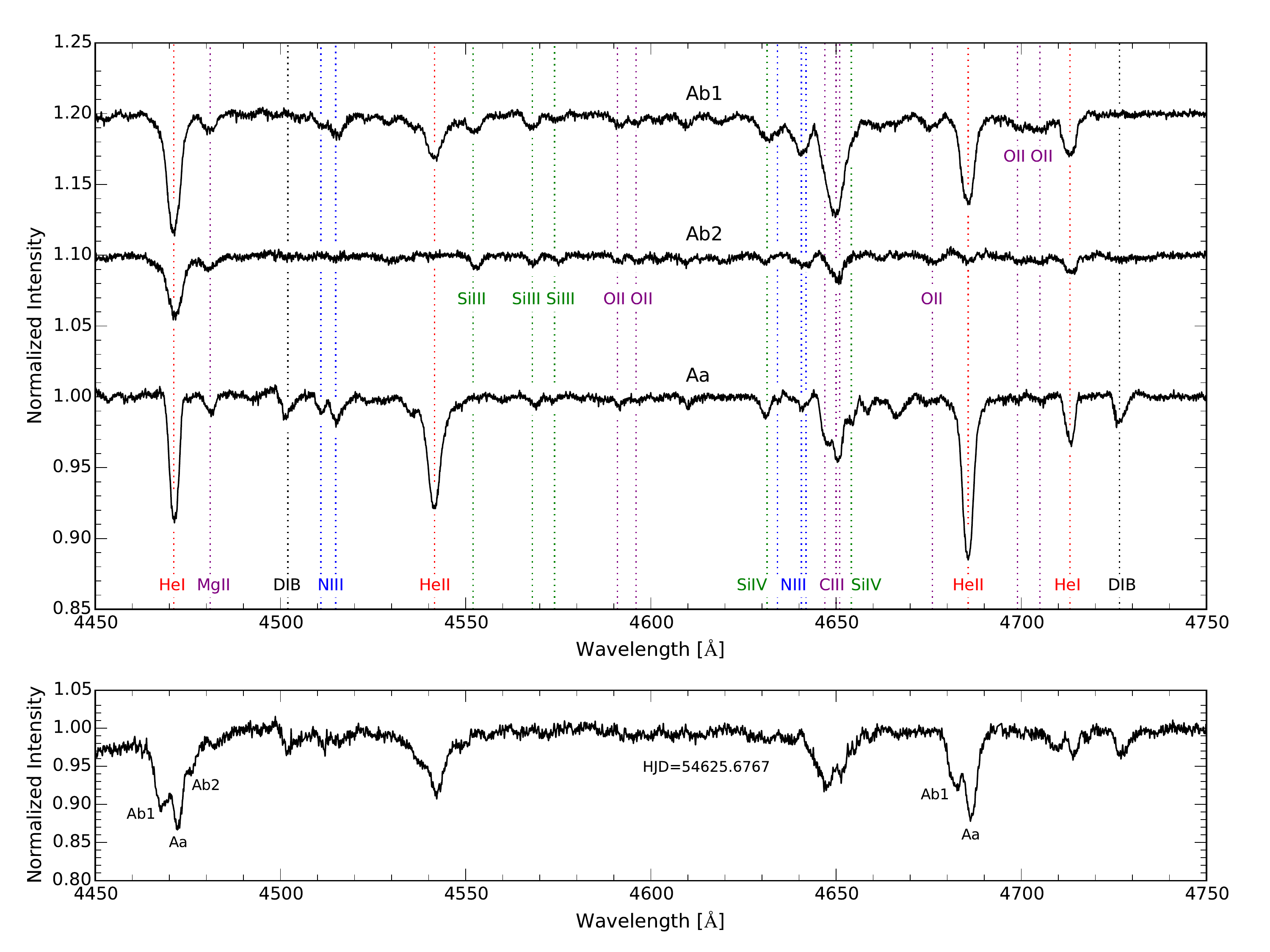}
\caption{Upper panel: normalized disentangled spectra of the three stellar components of H36A, in the range $4450-4750$ \AA.  Important spectral lines are identified as well as several diffuse interstellar bands (DIB). Data have been scaled to account for the dilution effect. Lower panel: composite spectrum of H36A obtained with FEROS. The spectrum is labeled with the corresponding Ab1, Ab2 and Aa component.
}
\label{fdisent}
\end{figure*}

In order to work with a homogeneous sample, and because the spectra obtained with the FEROS spectrograph have a larger resolution, we applied the method to this data subset only. 
We chose the $3900-5000$~\AA\,spectral range as it contains most of the features suitable for the spectral classification and analysis of massive stars.  
The convergence was achieved in 10 block-iterations.
Figure \ref{fdisent} shows a portion of the reconstructed spectra for the three stellar components. 
RVs and their errors obtained from the application of the method to the FEROS spectra are presented in Table~\ref{RVFEROS}.
For the \'echelle spectra from LCO, RVs were derived from the \ion{He}{ii}~$\lambda4686$ absorption line by fitting its profile with a sum of two Gaussian functions, one for each of the O-type stellar components (the B-type component does not contribute to this profile, see Sec.~\ref{specclass} for details on the spectral classification). 
Amplitude and dispersion for the two Gaussians were fixed in base of the values determined for the two O stars in the disentangled spectra.
The measurements corresponding to the LCO spectra used in the computation of the RV solution, i.e. the 21 best in terms of orbital phase and S/N, are listed in Table~\ref{RVLCO}.

Since the continuum flux in the spectrum of H36A is the sum of the contributions of three stellar components, the spectral features in the individual disentangled spectra appear "diluted".
Dilution factors were determined in an iterative way starting from the assumption that the binary and the third component contribute with roughly equal amounts of energy to the total flux, in agreement with the best-fit flux ratio 
$f(\textrm{\Cp} + \textrm{\Cs}) / f(\textrm{\Ct}) = 0.95 \pm 0.12$ found by \citet{sanchez14} for the combined $H$ and $K$-bands. 
A second starting assumption involves the flux ratio of the inner binary components which, according to their preliminary spectral types, was chosen to be $f(\textrm{\Cs})/f(\textrm{\Cp}) = 0.5$.
The iterative method consists of three steps. First, each of the disentangled spectra is multiplied by a suitable factor to correct the dilution. 
Then, the three individual corrected spectra (after being shifted according to the corresponding RV) are combined into a composite spectrum. 
Finally, the latter is checked against true spectroscopic data, and the values of the dilution factors are adjusted from the visual comparison. 
The process is repeated until a satisfactory match is achieved. 
As a result, it was found that the Aa, Ab1, and Ab2 components contribute to the global blue spectrum by 46\%, 38\%, and 16\% (with an uncertainty of about 2\%), respectively. 
The S/N of the final spectra is 280 for Aa, 230 for Ab1, and 130 for Ab2.

\begin{table*}
  \centering
  \caption{Radial velocities and their errors derived from the application of the spectral disentangling method to the FEROS spectra.}
  \begin{tabular}{rc rr rr rr}
\hline
\hline
N&  HJD($-$2400000)& VR$_{\text{\Ct}}$& \E$_{\text{\Ct}}$& VR$_{\text{\Cp}}$& \E$_{\text{\Cp}}$& VR$_{\text{\Cs}}$& \E$_{\text{\Cs}}$\\
\hline
 1&   53965.5510&    27.6&  2.0&    $-$18.4&   3.4&        37.5&  10.5\\
 2&   53967.5807&     5.9&  2.1&   $-$156.4&   2.8&       268.7&   9.2\\
 3&   54210.8726&    37.0&  1.1&   $-$154.7&   3.3&       225.6&   6.2\\
 4&   54210.9020&    36.1&  1.3&   $-$168.8&   3.3&       241.3&   6.2\\
 5&   54211.8558&    34.3&  1.4&      159.1&   2.6&    $-$302.4&   7.1\\
 6&   54246.6554&     1.6&  1.5&   $-$122.3&   2.9&       237.8&   5.3\\
 7&   54246.8581&    18.6&  2.5&    $-$15.6&   4.8&        39.8&  14.9\\
 8&   54247.7229&     7.5&  1.7&    $-$90.0&   3.0&       127.7&   7.7\\
 9&   54247.8960&     3.2&  1.4&   $-$146.7&   2.6&       269.3&   6.6\\
10&   54599.7856&    51.1&  1.3&   $-$103.4&   2.2&        98.7&   5.3\\
11&   54600.7773&    50.8&  1.3&   $-$107.0&   2.5&       109.6&   5.0\\
12&   54601.7546&    49.6&  1.6&      150.0&   2.4&    $-$313.2&   5.1\\
13&   54601.8789&    51.4&  1.4&      150.3&   2.7&    $-$311.3&   6.4\\
14&   54625.6767&    54.1&  1.1&   $-$206.9&   2.8&       263.2&   4.0\\
15&   54625.7851&    51.4&  1.2&   $-$201.1&   2.4&       257.9&   3.7\\
16&   54625.8946&    53.1&  1.7&   $-$159.3&   4.4&       189.0&   5.1\\
17&   54626.6375&    53.0&  1.9&      123.7&   3.2&    $-$267.6&   8.2\\
18&   54626.8825&    50.2&  1.0&    $-$36.7&   2.3&      $-$8.2&   4.8\\
19&   54627.6221&    51.3&  1.7&    $-$38.1&   2.8&      $-$7.4&  10.3\\
20&   54954.8244&     8.8&  2.2&      186.2&   2.5&    $-$287.6&   8.0\\
21&   54955.6602&     8.1&  1.6&   $-$176.4&   2.9&       292.5&   6.3\\
22&   54955.8749&     6.7&  1.7&    $-$85.2&   1.9&       157.2&   6.1\\
23&   54956.8478&    11.7&  2.0&    $-$53.9&   3.2&       103.5&   7.8\\
24&   57116.7975&    52.6&  1.7&   $-$193.2&   3.6&       244.3&   6.5\\
\hline
\hline
  \end{tabular}
  \label{RVFEROS}
\end{table*}

\begin{table}
  \centering
  \caption{Radial velocities  derived from the \ion{He}{ii}~$\lambda4686$ absorption line  in LCO spectra.}
  \begin{tabular}{rc rr rr}
\hline
\hline
N&  HJD($-$2400000)& VR$_{\text{\Ct}}$& \E$_{\text{\Ct}}$& VR$_{\text{\Cp}}$& \E$_{\text{\Cp}}$\\
\hline
  1&   53873.9166&    $-$56.7&      4.1&    126.0&      2.6\\
  2&   53875.9221&    $-$52.4&      1.7&    182.7&      2.5\\
  3&   53920.6423&    $-$28.2&      3.8&    151.2&      1.6\\
  4&   53936.6823&    $-$11.4&      1.2&   $-$161.6&      1.9\\
  5&   53937.6292&    $-$11.9&      1.2&    126.4&      1.3\\
  6&   54258.8971&     $-$3.7&      1.2&   $-$160.3&      2.9\\
  7&   54258.9087&     $-$3.6&      1.2&   $-$161.6&      1.7\\
  8&   54670.6413&     44.3&      1.6&   $-$129.4&      2.4\\
  9&   54670.7486&     44.0&      1.6&    $-$57.3&      2.0\\
 10&   54671.6735&     44.8&      2.1&    $-$97.8&      2.0\\
 11&   54671.7801&     44.2&      1.5&   $-$153.3&      2.3\\
 12&   56498.6207&     37.1&      3.6&   $-$178.2&      2.3\\
 13&   56498.6445&     37.2&      3.9&   $-$183.9&      2.1\\
 14&   56498.6678&     37.0&      3.4&   $-$193.5&      2.0\\
 15&   56813.8331&    $-$68.3&      3.9&    207.0&      1.5\\
 16&   56813.8564&    $-$67.8&      2.6&    215.6&      2.5\\
 17&   56813.8796&    $-$67.9&      2.9&    223.1&      2.3\\
 18&   56816.8360&    $-$68.4&      3.8&    173.7&      2.5\\
 19&   56816.8596&    $-$67.9&      2.8&    182.9&      2.3\\
 20&   56816.8833&    $-$68.8&      1.2&    198.3&      2.6\\
 21&   57570.7970&     51.6&      1.4&    142.3&      2.3\\
\hline
\hline
  \end{tabular}
    \label{RVLCO}
\end{table}

\subsection{Spectral classification}
\label{specclass}

The disentangled spectra (corrected by the appropriate dilution factors; see Sec.~\ref{separation}) were used to fine-tune the spectral classification of the stellar components in the triple system. 
This was performed  according to the Morgan–Keenan (MK) process, i.e., by %direct 
comparison of each individual spectrum with a grid of classification standard stars.
In order to use the standards from the {\em Galactic O-Star Spectroscopic Survey} (GOSSS; \citealt{sota11,sota14,maiz16}), the high-resolution disentangled spectra were degraded to the typical resolving power of the GOSSS data ($R\sim 2500$). The comparison between the H36A spectra and the standards was performed by direct visual inspection and also using \texttt{MGB} \citep{maiz12,maiz15b}, an IDL code specially developed for this task. 

\begin{figure*}
\centering
 \includegraphics[width=1.0\textwidth]{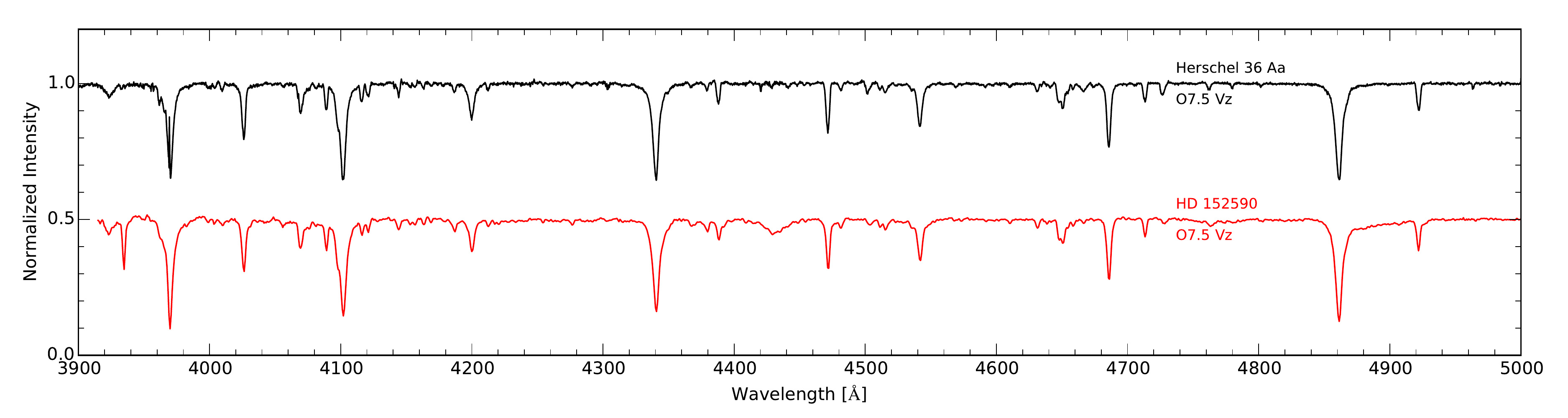}\\
 \includegraphics[width=1.0\textwidth]{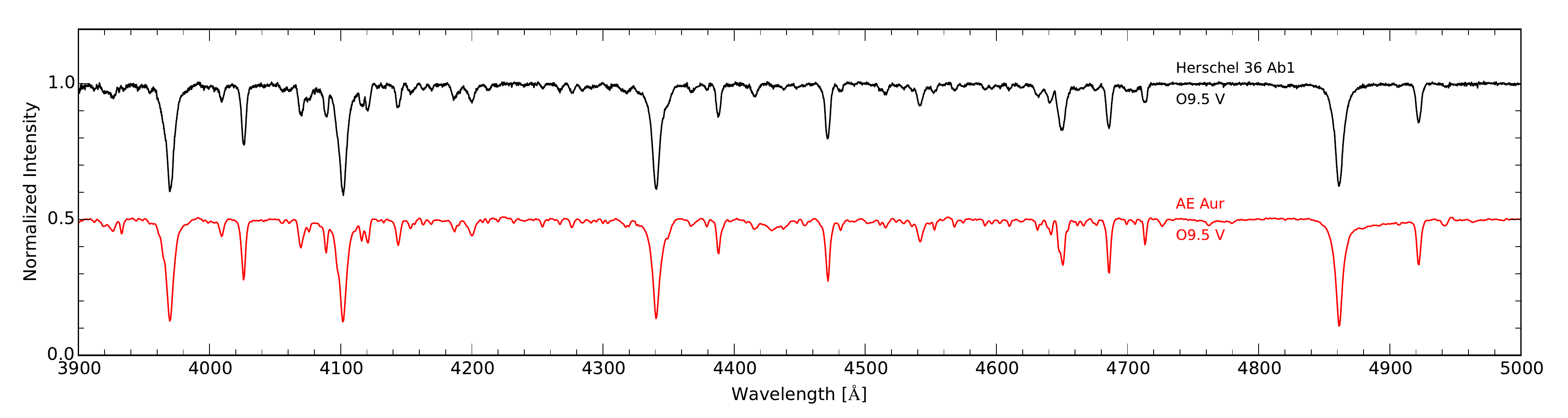}\\
 \includegraphics[width=1.0\textwidth]{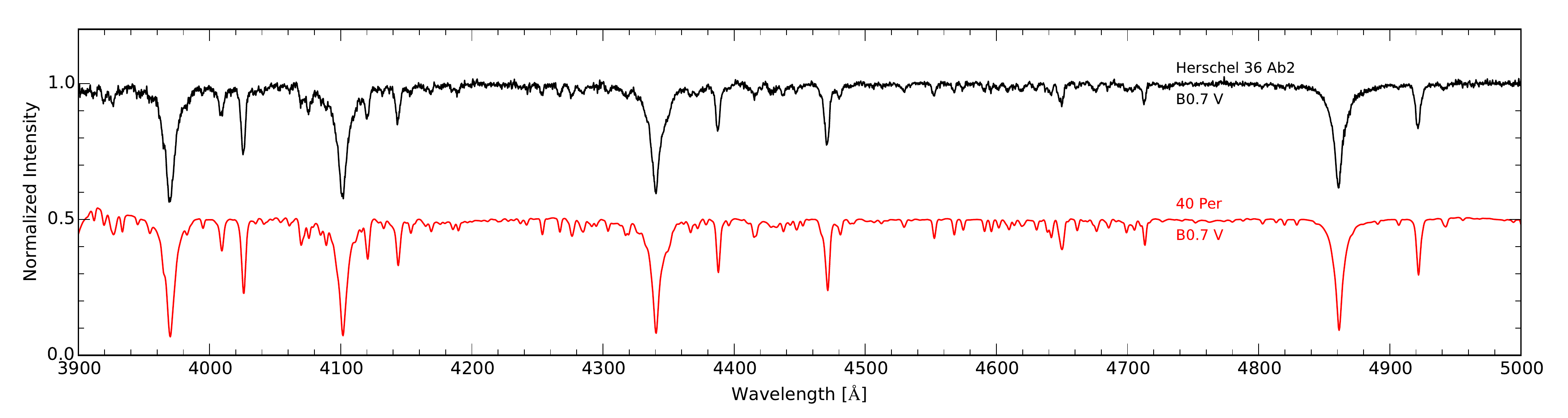}
 \caption{Spectral classification of the stellar components of H36A. Top/middle/bottom panel shows the individual spectrum of the components  Aa, Ab1, and Ab2 (in black) along with the MK classification standard stars (in red) corresponding to the assigned spectral type. The spectra of the standard stars are from the GOSSS database. %\citet{sota11} and \citet{maiz16}. 
}
\label{fclass}
\end{figure*}

Figure~\ref{fclass} shows the scaled spectrum of each component together with that of the standard star that better represents its spectral type. 
What follows is a brief description of the spectral classification differences with respect to the previous work by \citet{arias10}.

{\em Herschel~36 Aa}. The spectral type assigned to this star is O7.5\,Vz, which means the addition of the "z" qualifier to the previously determined luminosity class. 
The "z" qualifier (\citealt{walborn07}; see also \citealt{sota11}) refers to an unusually strong absorption in the \ion{He}{ii}~$\lambda4686$ line. The "OVz phenomenon", described in detail in \citet{sabin14} and \citet{arias16}, is hypothesized to be caused by the extreme youth of the objects, although additional factors may also play a role.
From the introduction of OVz standards in the grid \citep{maiz16}, the "z" qualifier is assigned through the classical MK methodology of visual comparison. 
However a quantitative criterion proposed by \cite{arias16} can give a measure of the "magnitude" of the OVz phenomenon. 
The criterion, applicable to spectral types earlier than O8.5, states that an O-type spectrum is classified as Vz if the $z$ parameter, defined as the ratio of the equivalent width of the  \ion{He}{ii}~$\lambda4686$ line to the maximum between the equivalent widths of the \ion{He}{i}~$\lambda4471$ and \ion{He}{ii}~$\lambda4542$, is greater or equal than 1.1. For H36Aa, we obtain $z=1.5$, which is one of the largest values derived so far for this parameter.

{\em Herschel~36 Ab1}. The spectral type assigned to this star is O9.5\,V, based on the absorption line ratios \ion{He}{ii}~$\lambda4200$/\ion{He}{i}~$\lambda4144$ and 
\ion{He}{ii}~$\lambda4542$/\ion{He}{i}~$\lambda4387$, which are smaller than unity (see \citealt{sota11} for details on the spectral classification criteria). 
This is two subtypes cooler than the previous determination. 

{\em Herschel~36 Ab2}. The published grid of classification standards of the GOSSS project includes types earlier than B0. 
Therefore, for the classification of this star we considered the standards and criteria described in \cite{walborn90}. 
At the same time we used a still unpublished version of MGB whose grid contains B-type standards.
From the ratio of the \ion{Si}{iii}~$\lambda4552$ to the \ion{Si}{iv}~$\lambda4089$ absorptions, which is slightly larger than unity, we assigned a B0.7\,V spectral type, i.e., one subtype cooler than the previous determination.

%########################################################### 
\subsection{Orbital solutions}
\label{orbits}

As shown by \cite{arias10}, Herschel~36\,A is a hierarchical system composed of a close binary in wide orbit around a third star. To compute the orbital elements %and their errors 
that characterize this triple system we used the IDL code 
\texttt{orbit3.pro}\footnote{The code is posted at \url{http://doi.org/10.5281/zenodo.321854}.} developed by Andrei Tokovinin. 
This is an interactive program for the calculation of visual, spectroscopic, or combined visual-spectroscopic orbits. 
For a triple system, it allows to fit the inner and outer orbits simultaneously. 

\citet{tokovinin17} explain in detail how \texttt{orbit3.pro} describes the triple system through 20 orbital elements, 10 for the inner pair (in our case, Ab1 + Ab2), and 10 for the outer pair (Aa + Ab, where Ab denotes the center of mass of Ab1 + Ab2).
As the center of mass of the inner pair, Ab, moves in the outer orbit, the RVs of Ab1 and Ab2 are sums of the inner and outer orbital velocities. 
On the contrary, the RV of Aa depends only on the outer elements. 
In general terms, the 20 orbital elements are required as input and then corrected iteratively. However, some of them are irrelevant for the fit in the case, like the present one, of a spectroscopic triple system for which the visual orbit is unknown. 
These elements are: the semi-major axes, the position angles of the line of the nodes, the orbital inclinations, and the "wobble factor", defined as the ratio of the astrometric amplitude to the semi-major axis of the inner orbit (see \citealt{tokovinin17}).

\begin{figure}
\centering
\includegraphics[width=1.0\columnwidth]{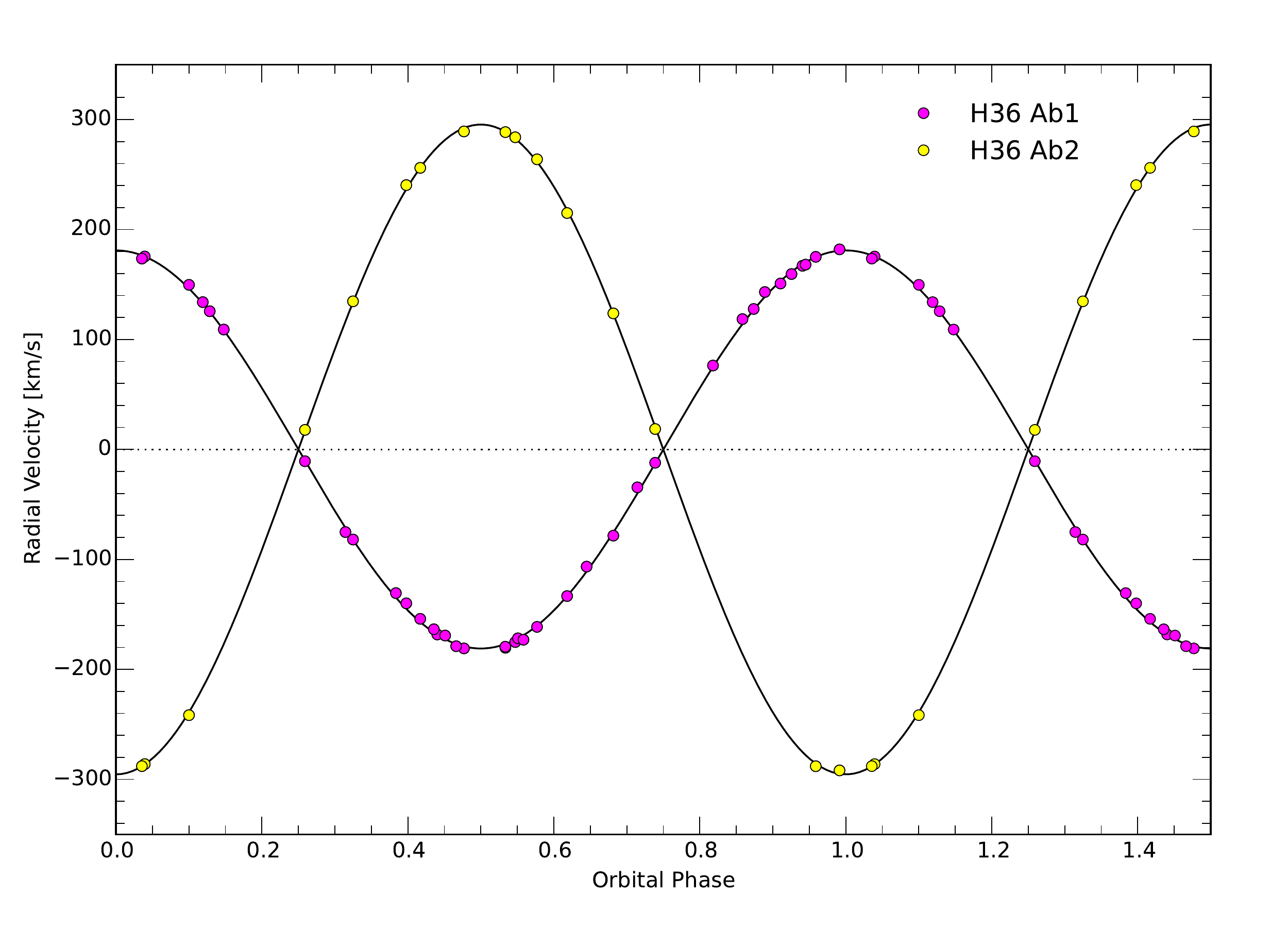}
\caption{RV curves corresponding to the inner binary Ab1+Ab2. %derived with \texttt{orbit3.pro} code. 
The RVs of each component have been corrected by the motion of the barycenter in the outer orbit.}
\label{fpar}
\end{figure}

\begin{figure}
\centering
\includegraphics[width=1.0\columnwidth]{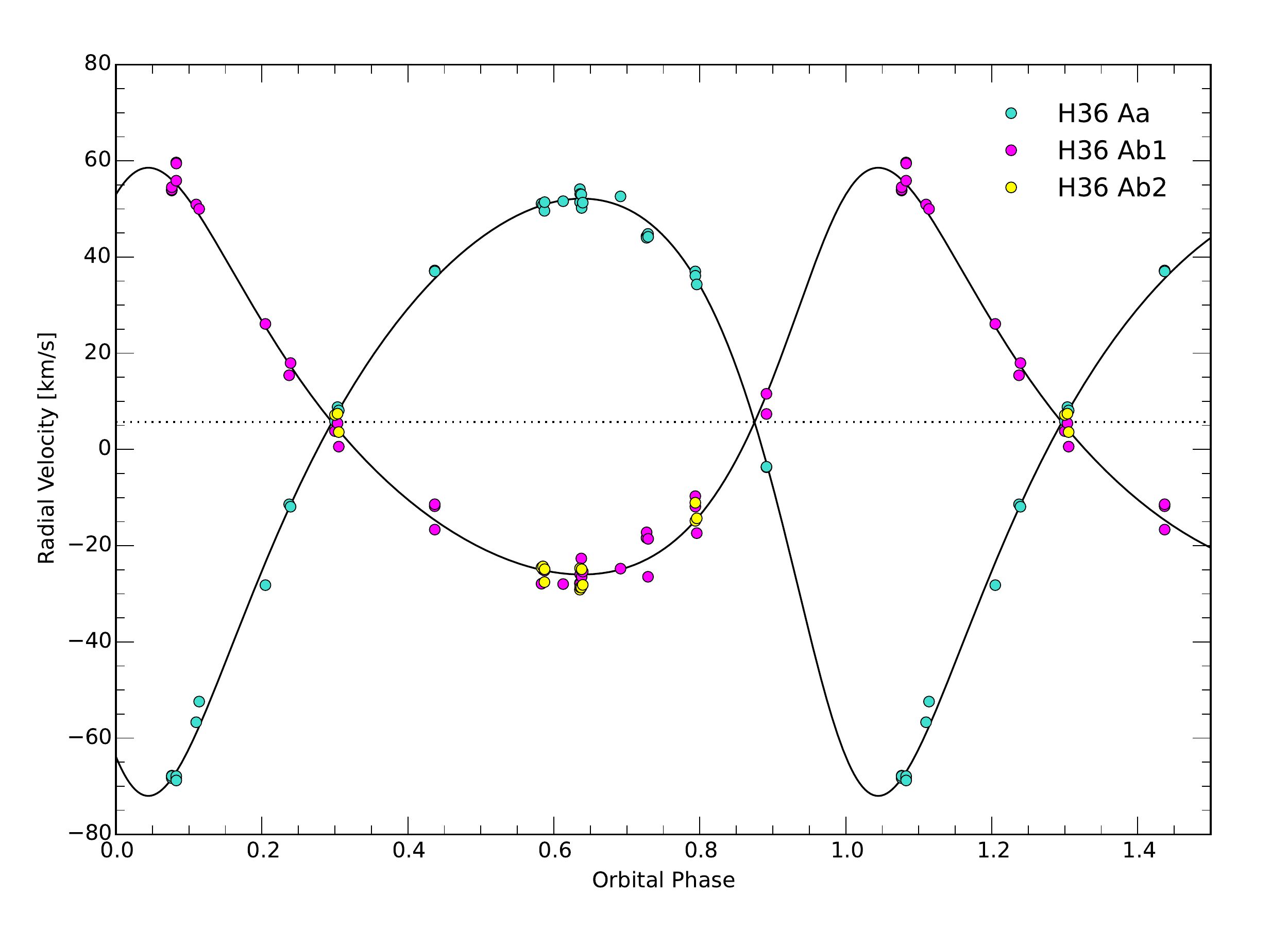}
\caption{RV curves corresponding to the outer orbit between the star Aa and the barycenter of the inner pair Ab1+Ab2.}
\label{fsys}
\end{figure}

The inner and outer orbital elements and their errors determined with \texttt{orbit3.pro} are presented in Tables \ref{tpar} and \ref{tsys}, respectively. 
The elements listed in Table~\ref{tpar} are: 
the inner orbital period ($P_{\rm in}$), 
the epoch of maximum RV for the component Ab1 ($T_{\rm RV_{\rm max}}$), 
the eccentricity of the inner orbit ($e_{\rm in}$, fixed), 
the RV semi-amplitudes $K$ for the components in the inner orbit, 
the semi-axes $a \sin i$ projected according to the inclination of the inner orbit $i_\textrm{in}$ , 
the minimum masses $M \sin^3 i$ of each component, 
and the mass ratio $q$ of the inner binary.
The elements listed in Table~\ref{tsys} are: 
the outer orbital period ($P_{\rm out}$), 
the periastron passage ($T_0$), 
the eccentricity of the outer orbit ($e_{\rm out}$), 
the argument of periastron ($\omega$), 
the systemic velocity ($V_0$), 
the RV semi-amplitudes $K$ for the component Aa and the barycenter of the pair Ab1 + Ab2, 
the semi-axes $a \sin i$ projected according to the inclination of the outer orbit $i_\textrm{out}$ (note that inner and outer orbits can have different inclinations), 
the minimum masses $M \sin^3 i$ of the component Aa and the binary Ab, 
and the mass ratio $q$ of the outer system. 
The rms values of the orbital solutions are also quoted in the last line of each Table. 
The eccentricity of the inner orbit was fixed at zero (left as a free parameter, it converges to a value smaller than 0.007 with an error comparable in magnitude).
The corresponding RV curves and solutions are shown in Figures \ref{fpar} and \ref{fsys}.

From all our data we selected those which had better S/N and were obtained at orbital phases more convenient to determine RVs for the system's components. 
This procedure left us with 38 spectra to be used in the orbital determination.
Their RV measurements, %(17 from ESO and 21 from LCO)
along with their residuals with respect to the orbits presented in the previous paragraph, are shown in Table \ref{tvrall}. 
In that table, the meaning of columns is as follows: 
(1) Heliocentric Julian Day of the observation; 
(2) observed RVs for the component Aa, V(Aa); 
(3) observed minus calculated, (O-C), values of V(Aa) in the outer orbit; 
(4) RV of the barycenter of Ab in the outer orbit VBC(Ab1) derived from the orbit of Ab1; 
(5) (O-C) values of VBC(Ab1); 
(6) RV of the barycenter of Ab in the outer orbit VBC(Ab2) derived from the orbit of Ab2; 
(7) (O-C) values of VBC(Ab2); 
(8) observed RVs for component Ab1;
(9) (O-C) values for component Ab1 in the inner orbit, where the observed values are the RV corrected by the barycenter motion in the outer orbit;
(10) observed RVs for component Ab2;
(11) the same as column (9) but for component Ab2.

\begin{table}
\begin{center}
\caption{Orbital elements for the inner binary Ab1 + Ab2 determined with \texttt{orbit3.pro} code. 
A circular orbit is assumed. 
The notation $i_{in}$ refers to the inclination of the inner orbit of this hierarchical triple system.}
\begin{tabular}{l r r}
  \hline
  \hline
  Parameter                                           & Value        & $\sigma$ \\
  \hline
  $P_{\rm in}$ (d)                                             & 1.54157      & 0.00006 \\
  $T_{\rm RV_{\rm max}}$ (HJD)                                       & 2,454,954.84 & 0.03    \\
  $e_{\rm in}$                                                 & 0.0          &         \\
  $K_{\text{\Cp}}$ (\kms)                             & 181.2        & 0.7     \\
  $K_{\text{\Cs}}$ (\kms)                             & 295.4        & 1.7     \\
  $a_{\text{\Cp}} \sin i_{\text{in}}$ ($10^5$ km)     & 38.40        & 0.15    \\  
  $a_{\text{\Cs}} \sin i_{\text{in}}$ ($10^5$ km)     & 62.62        & 0.36    \\ 
  $M_{\text{\Cp}} \sin^3 i_{\text{in}}$ (M$_{\odot}$) & 10.7         & 0.1     \\
  $M_{\text{\Cs}} \sin^3 i_{\text{in}}$ (M$_{\odot}$) & 6.6          & 0.1     \\
  $q_{(\text{\Cs}/\text{\Cp})}$                       & 0.613        & 0.006   \\
  rms (kms)                                           & 2.5          &         \\
  \hline
  \hline
\end{tabular}
\label{tpar}
\end{center}
\end{table}

\begin{table}
\begin{center}
\caption{
Orbital elements for the pair Aa + Ab, where Ab denotes the barycenter of the inner binary, determined with \texttt{orbit3.pro} code. 
The notation $i_{out}$ refers to the inclination of the outer orbit of this hierarchical triple system.}
\begin{tabular}{l r r}
  \hline
  \hline
Parameter&      Value&      $\sigma$ \\
\hline
  $P_{\rm out}$  (d)                                             & 492.8     & 0.7   \\
  $T_{0}$ (HJD)                                        & 2,454,805 & 17    \\
  $e_{\rm out}$                                                  & 0.29      & 0.01  \\
  $\omega_{\rm out}$ ($^\circ$)                                  & 330.6     & 0.9   \\
  $V_0$    (\kms)                                      & 5.7       & 0.4   \\
  $K_{\text{\Ct}}$  (\kms)                             & 62.0      & 0.6   \\
  $K_{\text{Ab}}$   (\kms)                             & 42.4      & 0.8   \\
  $a_{\text{\Ct}} \sin i_{\text{out}}$ ($10^6$ km)   & 402.1     & 5.7   \\
  $a_{\text{Ab}} \sin i_{\text{out}}$  ($10^6$ km)   & 275.3     & 6.4   \\
  $M_{\text{\Ct}} \sin^3 i_{\text{out}}$ (M$_{\odot}$) & 20.6      & 0.8   \\
  $M_{\text{Ab}} \sin^3 i_{\text{out}}$  (M$_{\odot}$) & 30.2      & 1.1   \\
  $q_{(\text{\Ct}/\text{Ab})}$                         & 0.681     & 0.072 \\
  rms (\kms)                                           & 2.0       &       \\
\hline
\hline
\end{tabular}
\label{tsys}
\end{center}
\end{table}

\section{Discussion}
\label{discussion}

\subsection{Stability and masses}

According to the orbital elements obtained in Sec.~\ref{orbits}, the inner binary revolves in a very tight ($P_{\rm in} = 1.54157$~d) circular orbit, 
at the time that moves in an eccentric ($e=0.29$), much wider ($P_{\rm out} = 492.8$~d) orbit 
around the center of mass of the triple system. 
According to \citet{mardling01}, a triple system is said to be stable if the periastron separation of the outer binary  ($R_{\rm out}$) satisfies the condition

\begin{equation}
		R_{\rm out} > 2.8\ \Bigg[ (1+q_{\rm out}) \frac{1+e_{\rm out}}{(1-e_{\rm out})^{1/2}} \Bigg]^{2/5}  a_{\rm in},
		\label{stability}
	\end{equation}
	
\noindent where $a_{\rm in}$ is the major semi-axis of the inner orbit. For H36A, this relation is widely satisfied as 	$R_{\rm out}$ is approximately 100 times the minimum value for stability. The system is thus within the hierarchical space for triples in which the inner component is barely affected by perturbations from the third star.

%The ratio of outer to inner period $P_{\rm out}/P_{\rm in}$  
%satisfies the "empirical stability criterion" by \citet{tokovinin04} $P_{\rm out} (1 - e_{\rm out})^3 P_{\rm in} > 5$, placing the system well within the stability zone for hierarchical triples.

The spectroscopic mass ratio obtained for the inner binary ($q_{\rm in}=0.613$) points to an appreciable mass difference ($\sim$38$\%$) between the stellar components Ab1 and Ab2. 
On the other hand, the mass ratio derived from the outer orbit ($q_{\rm out}=0.681$) implies that the mass of the star Aa is $\sim$2/3 of the total mass of the inner binary.

Since we lack direct information about the inclinations of the orbits, we are unable to derive absolute masses.
However, further information can be obtained by combining the inner and outer orbital solutions. 
From $M_{\rm Ab} \sin^3 i_{\rm out}=30.2$~M$_{\odot}$ and the mass ratio $q_{\rm in}=0.613$, it turns that $M_{\rm Ab1} \sin^3 i_{\rm out}=18.7$~M$_{\odot}$. 
Even under the assumption of $i_{\rm out}=90^{\circ}$, this value exceeds that expected for an O9.5 dwarf according to the current theoretical calibrations. 
For example, the spectroscopic mass derived by \citet{martins05} from the calibration of the effective temperature of O stars is 16.5~M$_{\odot}$ for this spectral type. 
Now, if $i_{\rm out}\approx90^{\circ}$, the mass of the most massive component turns to be $M_{\rm Aa} \approx 20.6$~M$_{\odot}$, which is lower than the mass expected for an O7.5\,V star (from the same calibration, $M=24.1~M_{\odot}$ for an O7.5\,V). 
Conversely, if the inclination of the outer orbit is such that $M_{\rm Aa}$ reaches the "expected" value of $24.1~M_{\odot}$ (i.e., $i_{\rm out} \approx 72^{\circ}$), then $M_{Ab1} \approx 21.7~M_{\odot}$. 
This simple exercise shows that it is impossible to conciliate our dynamical (model-independent) masses with those from the theoretical calibrations for both stars simultaneously. 
The mass difference between our O dwarfs of types O7.5 and O9.5 results substantially smaller than that predicted by the theory (less than 2.5~M$_{\odot}$; confront with the difference of 7.7~M$_{\odot}$ from Table~1 by \citealt{martins05}). 

Beyond the uncertainty introduced by the unknown orbital inclination, the minimum mass derived for the most luminous component, $M_{Aa}$ ($20.6$~M$_{\odot}$), is in agreement with other empirical determinations for non-evolved stars of similar spectral type. 
For example, \cite{rauw01} determined an absolute mass of $22.2$~M$_{\odot}$ for the main component of the eclipsing system CPD $-59^\circ$\,2603 (O7.5\,V(n)z, \citealt{sota14}). 
Another example is the "twin" eclipsing system EM~Car (O7.5 V((f)) + O7.5 V((f)), \citealt{maiz16}), for whose components \citet{andersen89} derived $21.4$ and $22.9$~M$_{\odot}$, respectively. 
Moreover, using the apsidal motion rate, \citet{ferrero13} calculated absolute masses of 22.5~M$_{\odot}$ and 20.5~M$_{\odot}$ for the primary and secondary of another twin binary, HD~165052 (O7.5\,Vz + O7.5\,Vz).

On the contrary, the minimum mass derived for the Ab1 component ($18.7$~M$_\odot$) is somewhat larger than expected from the comparison with members of eclipsing binaries of the same spectral type such as, for example, 
the secondary of CPD $-59^\circ$\,2603 (14.4~M$_{\odot}$, \citealt{rauw01}), 
the primary of CPD $-59^\circ$\,2628 (14.0~M$_{\odot}$, \citealt{freyhammer01}), 
V478~Cyg (15.3~M$_{\odot}$, \citealt{martins17}), 
the secondary of FO15 (16.0~M$_{\odot}$, \citealt{niemela06}), and the components of
HD~198846 (17.7~M$_{\odot}$, \citealt{burkholder97}).

\subsection{Relative inclination between inner and outer orbits}

We now explore the orbital "architecture" of the hierarchical triple system H36A. 
The Ab1+Ab2 pair moves in an orbit of inclination $i_{\rm in}$, while revolves in a larger orbit of inclination $i_{\rm out}$ around Aa. 
The values of $i_{\rm in}$ and $i_{\rm out}$ are constrained by the minimum masses derived from the spectroscopic orbital solutions.
As previously discussed, it is reasonable to assume $i_{\rm out} \approx 90^{\circ}$, such that $M_{Ab1}$ does not exceed too largely the typical mass of an O9.5 dwarf.
This subsequently leads to $i_{\rm in} \approx 56^{\circ}$, meaning that inner and outer orbits are not coplanar.
On the other hand, if $i_{\rm out} \approx 72^{\circ}$, such that $M_{Aa}$ resembles that expected according to the theoretical calibrations, 
the resulting value for the inner inclination is $i_{\rm in} \approx 52^{\circ}$. 
Whatever the case may be, the orbits are not coplanar.  The difference between the inclinations inferred for the orbital planes is in the range $[20^{\circ},34^{\circ}]$. However, since the orientation of each orbit on the sky is unknown, the former difference is only a lower limit on the relative orbital inclination. We thus conclude that the misalignment between both orbits is at least 20 degrees.

%The difference between their orbital inclinations provides a lower limit on the relative inclination between inner and outer orbits, which thus results to be larger than $20^{\circ}-34^{\circ}$.

Along with other parameters such as the masses and eccentricities, the orientation of the angular momenta of the inner and outer orbits in triple systems is closely related to their origin, and provides information about the dynamical and dissipative processes intervening in their formation (e.g. \citealt{bate14}, \citealt{antognini16}, \citealt{moe18}).
The alignment of the present orbits is connected to the alignment between the original circumstellar disks and the stellar spins (e.g. \citealt{wheelwright11}). In the case of a massive system, this in turn informs, for example, whether it forms via monolithic collapse and subsequent disk fragmentation (consistent with the model of massive star formation by \citealt{krumholz09}) or via stellar captures  (consistent with the formation of massive binaries proposed by \citealt{moeckel07}).

Unfortunately, massive triple systems where elements of both outer and inner orbits are known are very scarce. Most of the current statistics comes from the study of low-mass stellar systems (e.g. \citealt{tokovinin17st,tokovinin18st}).
While some triple systems have both orbits located in one plane, others show moderately to highly misaligned orbits. 
These misalignments are often accounted for by dynamical interactions with nearby members of the same cluster 
\citep{antognini16}.
\citet{tokovinin17st} shows that the orbit alignment depends on mass,   
being stronger in triple systems with low-mass primaries, compared to more massive stars. 
This means that chaotic stellar dynamics (e.g. collisions) may be important in the formation of massive multiple systems.
Misaligned triple systems may also be created  at that epoch of star formation when gas with randomly aligned angular momentum is accreted, changing the orientation of the outer orbit (\citealt{bate10}).

Like H36A, other massive hierarchical triple systems have the inner and outer orbits in different planes. 
An outstanding example is HD~150136, composed of an inner 2.67~d-period O3-3.5\,V((f*)) + O5.5-6\,V((f)) binary in a wider orbit ($\sim$3000~d-period) around an O6.5-7\,V((f)) star 
(\citealt{mahy12,sana13}).
Spectroscopic and astrometric observations point to a minimum relative inclination of $\sim$44$^{\circ}$ between both orbits 
\citep{mahy18}.
Another example is the $\sigma$ Orionis triple system 
(\citealt{sig_ori11,sig_ori15})
composed of an astrometric binary with a period of $\sim$156~years, whose primary component is at the same time a spectroscopic binary in a highly eccentric orbit with a period 400 times smaller ($P\sim143$~d). Recent interferometric observations allowed to resolve the orbit of the latter, confirming that inner and outer orbits are not coplanar  and have a relative inclination of $120^\circ$  or $126.6^\circ$ \citep{schaefer16}.
 Finally, it is worthwhile to mention the early-B multiple system HD~315031. Although somewhat less massive than H36A, this hierarchical triple system belongs to the same star forming region (it is a member of the young open cluster NGC~6530 in Messier~8), and shows similar orbital characteristics.  \citet{gonzalez14} studied in detail this system and concluded that it contains a very short-period ($1.37$-d) binary (B0.5\,IV-V + B1\,V) and a third star orbiting in an eccentric ($e=0.85$), much wider ($P=482.9$~d) orbit. From the stellar masses estimates, these authors found that the inclination between inner and outer orbits must be of $58^\circ$ or larger.

\subsection{Constraints from the astrometric observation}

Additional constraints to the total mass of the system can be derived from the AMBER interferometric observation obtained by \citet{sanchez14}, on the condition that a reliable estimate of the distance is available. 

For H36A, the GDR2 parallax is $\varpi = 0.9028 \pm 0.2192$~mas, which has a large relative uncertainty, leading to a biased result if one estimates the distance by simply inverting the parallax \citep{LutzKelk73,Maiz01a,luri18}. 
Furthermore, the quoted uncertainty in GDR2 is just the internal one (or $\sigma_{\rm int}$) and does not consider the effects of external random and systematic sources \citep{Lindetal18a,Lindetal18b}. 
Given those restrictions, to derive a GDR2 distance to Herschel 36 we take the following steps, where for the most part we use the strategy of \citet{Lindetal18b}:

\begin{itemize}
\item We start by selecting several O stars in M8 and calculating the distance to the O-type stars in the cluster instead of the distance to Herschel~36 only. The goodness of fit for GRD2 astrometry recommended by \citet{Lindetal18b} is the {\em Renormalized Unit Weight Error} (RUWE), a reduced $\chi^2$ corrected by a magnitude- and color-dependent factor. 
We calculated RUWE for five O-type systems in M8: Herschel~36, 9~Sgr~AB, HD~165052, HD~164816, and HD~164536. 
The value is lower than 1.4 for the first four targets and 2.22 for the last one. 
Therefore, we excluded it and used only the first four for the subsequent calculations.
\item We compute an external (or total) uncertainty $\sigma_e$ as:

   \begin{equation}
        \sigma_{\rm ext}^2 = k^2\sigma_{\rm int}^2 + \sigma_s^2,
   \end{equation}
       
\noindent with $k=1.08$ and $\sigma_s = 0.021$~mas, as all targets are in the bright-star regime of \citet{Lindetal18b}. 
\item We estimate the uncorrected (for possible parallax zero point) M8 parallax as:
 
    \begin{equation}
        \varpi_{\rm M8,u} = \sum_{i=1}^{4} w_i \varpi_i,
    \end{equation}

\noindent where:

    \begin{equation}
        w_i = \frac{1/\sigma_{{\rm ext,} i}^2}{\sum_{i=1}^{4} 1/
              \sigma_{{\rm ext,} i}^2}
    \end{equation}  

\noindent $\varpi_i$ are the weight and parallax of each of the selected four O-type systems, respectively.
\item The random uncertainty of the M8 parallax is calculated as:

    \begin{equation}
        \sigma_{\rm M8}^2 = \sum_{i=1}^{4} w_i^2 \sigma_{{\rm ext,} i}^2 + 
        2\sum_{i=1}^{3}\sum_{j=i+1}^{4} w_i w_j V_\varpi(\theta_{ij}), 
    \end{equation}

\noindent where $V_\varpi(\theta_{ij})$ is the parallax spatial covariance function for two objects separated by an angle $\theta_{ij}$. 
\item As described in \citet{Lindetal18b}, there is a significant parallax zero point present in GDR2 but it likely depends on position, magnitude, and color. 
What is known from an analysis of quasars that cover most of the sky is that the global average is $-0.030$~mas. As we currently cannot do better than that we simply calculate a corrected M8 parallax (in mas) as:

    \begin{equation}
         \varpi_{\rm M8,c} = \varpi_{\rm M8,u} + 0.030
    \end{equation}

\noindent and give the two values for the M8 parallax (corrected and uncorrected) as a measure of the systematic uncertainty in GDR2.
\item Finally, one has to consider the underlying spatial distribution of the sample to obtain a proper distance (a constant spatial distribution yields a singularity at zero parallax or infinite distance, see \citealt{LutzKelk73}). We use the method described by \citet{Maiz01a,Maiz05c}, appropriate for Galactic OB stars such as the ones considered here, assuming that they belong to the isothermal disk component described by the parameters calculated by \citet{Maizetal08a}.
\end{itemize}

Following the above procedure, we obtain an uncorrected M8 parallax of $0.813 \pm 0.044$~mas and a corrected one of $0.843 \pm 0.044$~mas. Note that the random and systematic errors are comparable. 
The associated distances are $1245^{+73}_{-65}$~pc (uncorrected) and $1199^{+67}_{-60}$~pc (corrected). 
Both distances are within one sigma of the value of 1250~pc derived by \citet{arias06} using the \texttt{CHORIZOS} code \citep{maiz04} to model the spectral energy distribution of the nearby stars and the extinction to the Hourglass Nebula. 
Therefore, we adopt the latter estimate for the distance to the system.

\begin{figure}
 \includegraphics[width=\columnwidth]{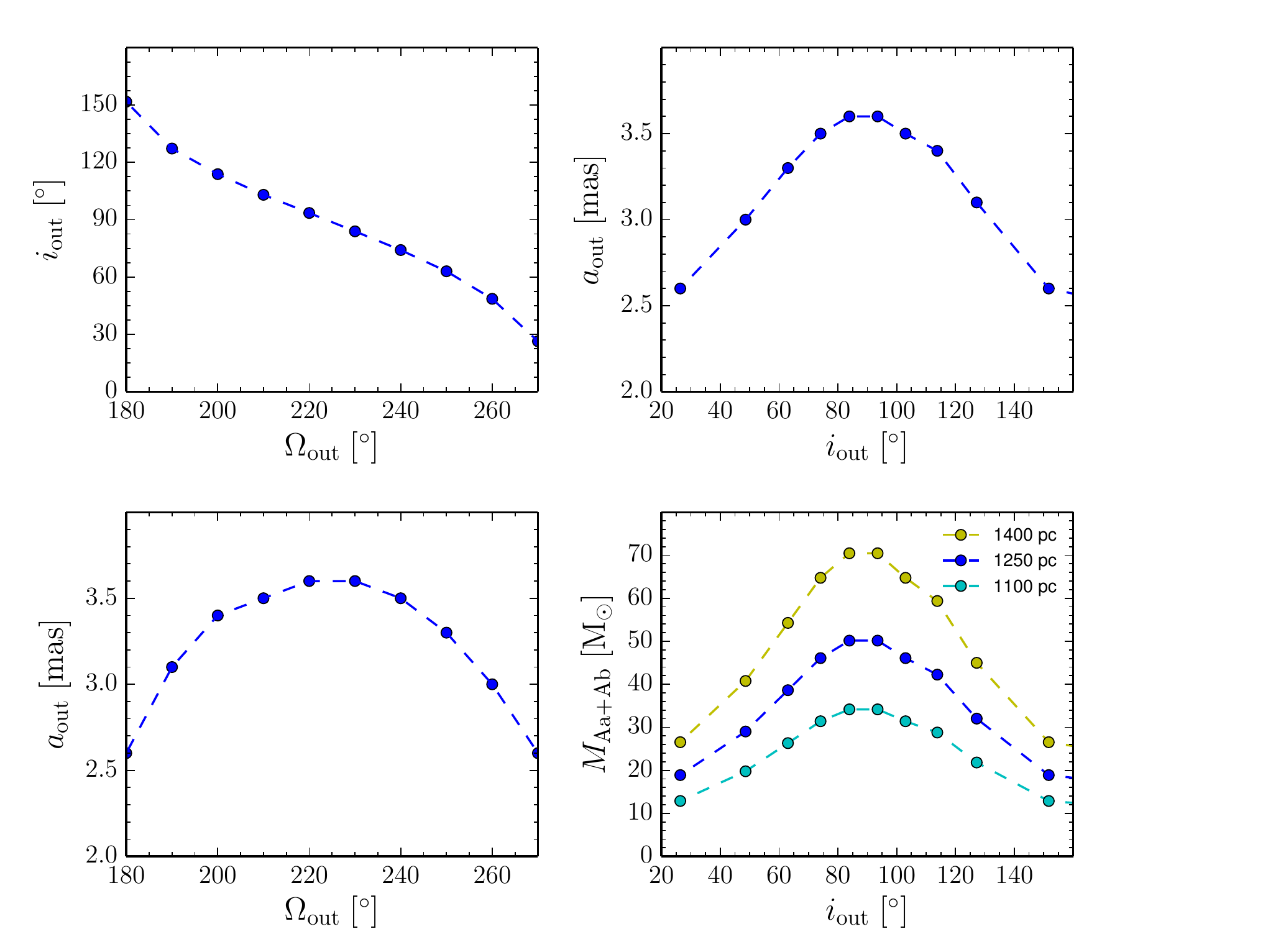}
 \caption{Analysis of the variation of the outer orbital elements with the position angle of the outer orbit line of the nodes, $\Omega_{\rm out}$. 
 Left-panels: variation of $i_{\rm out}$ (top) and $a''_{\rm out}$ (bottom) as a function of $\Omega_{\rm out}$. Top-right panel: variation of $i_{\rm out}$ as a function of $a''_{\rm out}$.
Bottom-right panel: total mass of the system ($M_{\rm Aa+Ab}$) calculated using the Equation~\ref{totalmass} as a function of $i_{\rm out}$ for three different values of the distance.}

\label{angle}
\end{figure}

The modelling of the AMBER observation yields an angular separation of $1.81\pm0.03$~mas between the component Aa and the binary Ab.
Once a distance is assumed ($d=1\,250$ pc in this case), we can infer the total mass of the system, $M_{\rm Aa+Ab}$, if we have information about the angular size of the outer orbit, $a''_{\rm out}$. 
From Kepler's third law,

   	\begin{equation}
		M_{\rm Aa+Ab} = \frac{4 \pi^2}{G} \frac{(d \cdot a''_{\rm out})^3}
        {P_{\rm out}^2}
		\label{totalmass}
	\end{equation}

\noindent where $G$ is the gravitational constant and $P_{\rm out}$ the period of the outer orbit.
Although only one astrometric datum is not sufficient to derive a robust value for the semi-major axis, we can put a constraint on that value making use of the obtained spectroscopic solution. 
To that aim, we run \texttt{orbit3.pro} anew but fixing all the orbital parameters except for the semimajor axis, the periastron passage and the inclination of the outer orbit, while varying only the position angle of the line of nodes ($\Omega_{\rm out}$).
The left-top panel of Figure~\ref{angle} shows that when the position angle $\Omega_{\rm out}$ varies between $180^\circ$ and $270^\circ$, a large range of orbital inclinations is in principle possible, specifically $26^\circ < i_{\rm out} < 151^\circ$.
However not all those inclinations are compatible with the minimum value of the total mass resulting from the spectroscopic solution, $M_{\rm Aa+Ab} \sin^3 i_{\rm out} = 50.8$~M$_\odot$ (see Table~\ref{tsys}). 
The right-bottom panel of the same Figure shows the total mass as a function of the inclination. 
To be in concordance with the spectroscopic solution, the adopted value of the distance ($1\,250$~pc, blue curve) implies  $i_{\rm out}\sim90^\circ$ and a maximum total mass of $M_{\rm Aa+Ab}=50.1$\,M$_\odot$.  

We considered two additional distance values ($1\,100$~pc and $1\,400$~pc, cyan and yellow curves in Figure~\ref{angle}, respectively) in order to check how the inferred total mass is sensitive to the adopted distance. 
We found that distances shorter than that adopted must be excluded since they yield total masses below the spectroscopic limit of $50.8$~M$_\odot$. 
On the other hand, distances larger than the adopted allow orbital inclinations different from $90^\circ$. 
For example, if the system was located at 1400~pc, and the total mass $M_{\rm Aa+Ab}$ was in the range $[50,60]$\,M$_\odot$, inclinations $i_{\rm out}$ in both the range $[58^\circ, 68^\circ]$ and  $[114^\circ, 124^\circ]$ could be possible.
As a conclusion, the requirement of a similar "astrometric" and "spectroscopic" total mass of roughly $50.8$~M$_\odot$ imposes a strong constraint in the distance to the system, which must be equal or larger than 1.25~kpc.

\subsection{Photometric information}

 Because the orbital inclinations estimated in our analysis suggest that stellar eclipses between the components of H36A might occur, we searched for the available photometric observations in the public databases. 

In the General Catalogue of Variable Stars (GCVS, \citealt{samus17}) H36A appears as a suspected variable (NSV~10081) based on the historic observations in the nineteenth century by Sir John Herschel \citep{herschel1847} and Johann Schmidt \citep{schmidt1868}. 
A catalogue of variable stars constructed from the "All-Sky Automated Survey for Supernovae" data (ASAS-SN, \citealt{shappee14}; \citealt{kochanek17}) has been very recently published by \citet{jayasinghe18}. H36A is included there 
as ASASSN-V~J180340.37-242241.3, a variable source of type Gamma Cassiopeia (GCAS) with a probable period of 371~days.  

We retrieved the photometric data from the ASAS-SN server\footnote{http://asas-sn.osu.edu}, which consist of a time-series of the $V$ or $g$-band magnitudes extracted in a fixed two-pixel (16'') aperture, with a background computed in an annulus of 7-10 pixel radius (56''-80''), and calibrated using the AAVSO
Photometric All-Sky Survey (APASS; \citealt{henden15}) catalog. 

The ASAS-SN images are obtained through a with of telescopes in both hemispheres. In the case of H36A, the images correspond to four different cameras which show a different behavior. The photometric time-series corresponding to the camera named "bh" resembles the observed in GCAS-type variables, with a minimum followed by a sharp rising. However, the behaviour of the "bg" camera data during the same period is  very different, showing scattered points around a mean value. Such a large difference between the photometry determined from both cameras casts doubt on 
the GCAS-variable type reported by \citet{jayasinghe18}. It must be noted that the area around H36A is very complex, showing several stellar sources and large changes in the brightness of the sky within few arc-seconds. In consequence, blending, crowding and background subtraction may be critically affecting the ASAS-SN results.

\begin{figure}
\includegraphics[width=\columnwidth]{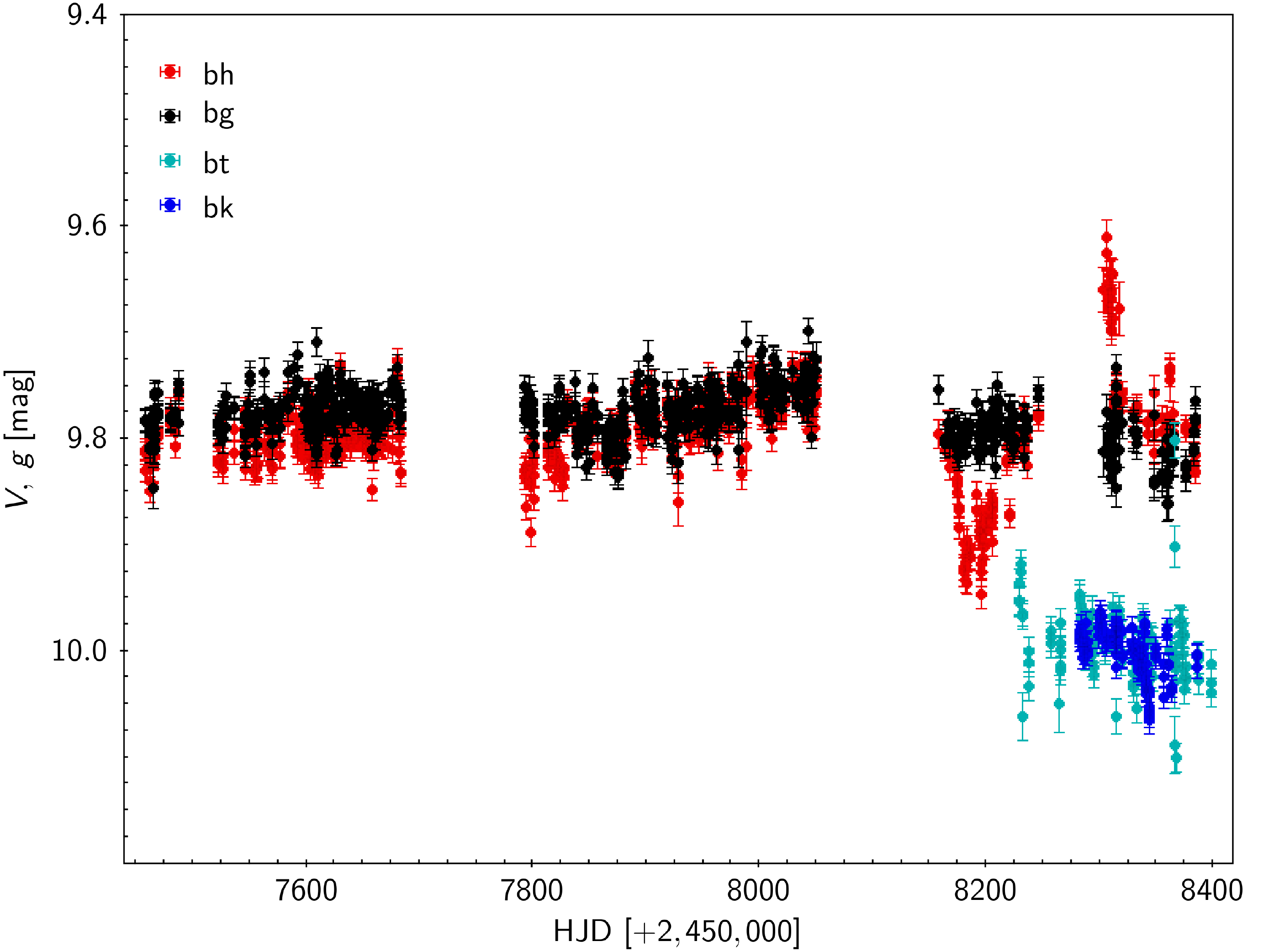}
\includegraphics[width=\columnwidth]{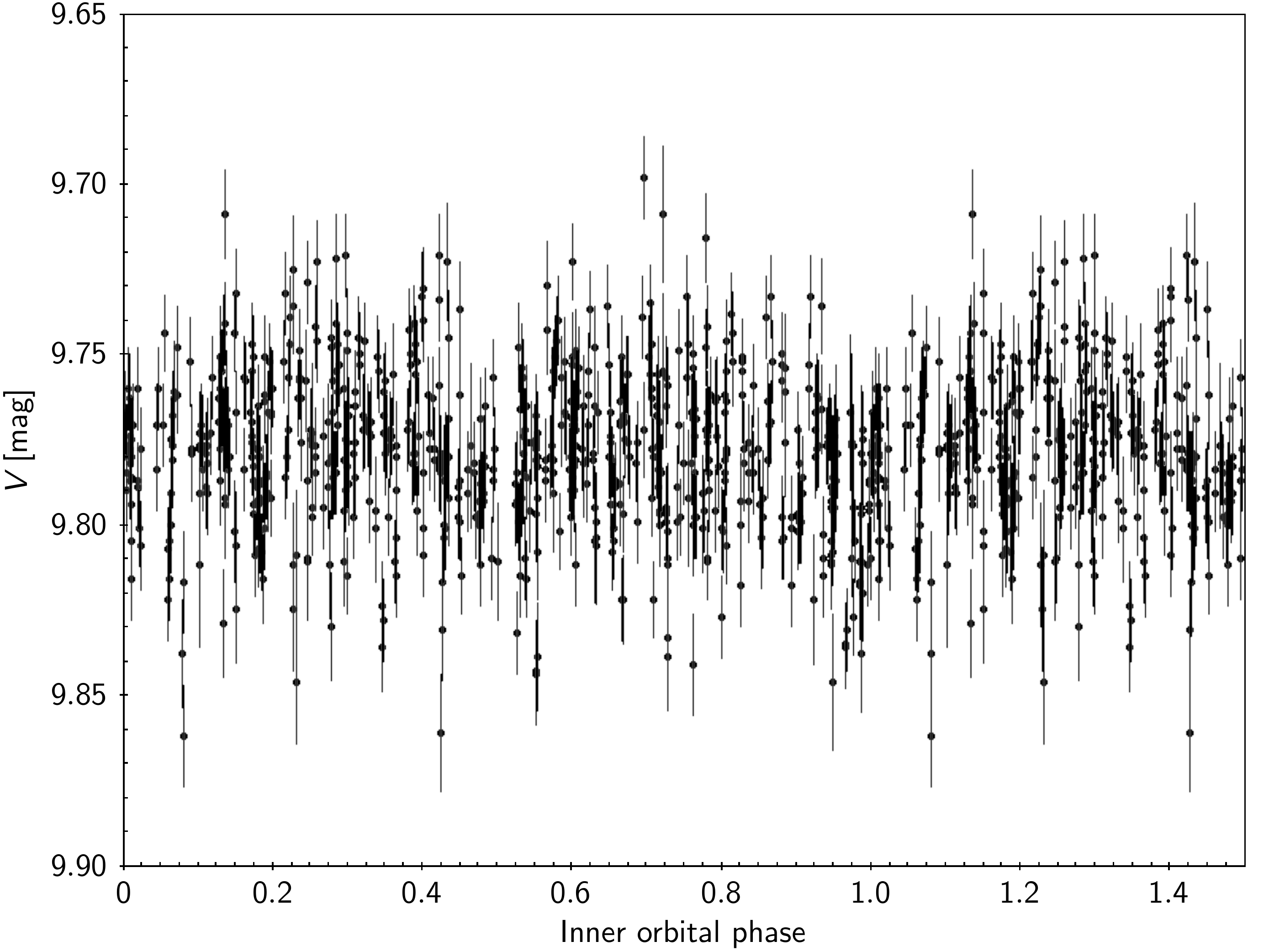}
\caption{Upper panel: ASAS-SN complete light-curve of H36A in $V$ and $g$ bands obtained with cameras "bg" (black), "bh" (red), "bk" (cyan) and "bt" (blue). Lower panel: light curve obtained with camera "bg" folded using the spectroscopic period of the inner orbit. The minimum at around phase=0.0 occurs during the orbital conjunction in which the primary O-star component is behind the B-star secondary.}
\label{asassn}
\end{figure}

On the other hand, we analyzed the ASAS-SN data corresponding to the camera "bg" 
and found that the period search 
brings a weak signal around 1.5417~d, a value very close to the spectroscopic orbital period of the inner binary. The bottom panel of Figure~\ref{asassn} 
shows the folded light curve constructed by phasing the ASAS-SN data against the spectroscopic period of 1.54157~d obtained in this work. The presence of shallow eclipses or ellipsoidal variations, is observed. One of the minima occurs during the orbital conjunction (phase 0.0) in which the primary O-star component is behind the B-star secondary. This featured light curve is in agreement with the orbital inclination of $\sim 55^\circ$ suggested for the close pair. 
The scenario resembles the triple massive system HD~150136 (\citealt{mahy18}), in which the inner binary (period $\sim2.67$~d) presents grazing eclipses, being in an orbit with an inclination of $62^\circ$. 
To summarize, although the quality of the ASAS-SN light curve does not allow further analysis, it provides considerably important evidence on the existence of stellar eclipses or ellipsoidal variations in H36A. Future photometric observations are essential. 
%but those open up the possibilities for more detailed observations. 

\section{Summary and outlook}
\label{outlook}

H36A has been previously shown to be a hierarchical triple, composed of a close binary (Ab1,Ab2) in wide orbit around a third star (Aa).  
In this paper we used a large set of high-resolution, high S/N optical spectroscopic observations to characterize the spectral and orbital properties of this massive system in the most accurate possible way. 
The disentangling of the composite spectrum allowed to improve the spectral classification of the individual components, which resulted in O7.5\,Vz, O9.5\,V and B0.7\,V for Aa, Ab1 and Ab2, respectively. 
High precision spectroscopic solutions were determined for the close binary orbit, the "inner" orbit, as well as for the orbit of Aa and Ab around the center of mass of the triple system, the "outer" orbit. 
Inner and outer orbits turned to be not coplanar, being their relative inclination of at least 20 degrees.
Dynamical minimum masses of $20.6$~M$_{\sun}$ for Aa, $18.7$~M$_{\sun}$ for Ab1, and $11.5$~M$_{\sun}$ for Ab2 were derived, in reasonable agreement with other empirical determinations and the current theoretical calibrations.  
We also used the only interferometric observation obtained so far to impose additional constraints on the total mass of the triple system as well as on its distance from the Sun. 
Our results point to a minimum total mass of $50.8$~M$_\odot$ and suggest the distance to the system must be equal or larger than 1.25~kpc.

The orbital properties of multiple stellar systems are closely related to their origin. 
Therefore, like other massive triple systems whose inner and outer orbits can be determined with reasonable accuracy, H36A is a key object to prove the current scenarios of massive star formation. 
High-resolution spectroscopic monitoring of massive stars, such as the OWN Survey, are crucial to find more interesting systems and determine their orbits. 
However, a complete orbital characterization cannot be obtained from RV-studies only. 
Either astrometric observations and/or photometric data, provided the system is eclipsing, are necessary. 

Since the orbital inclinations estimated in our analysis suggested the possibility of occurrence of stellar eclipses between the components of H36A, 
we investigated its photometric variability using the optical observations from the ASAS-SN public database and found evidence of shallow eclipses in the inner binary.  Further photometric observations are thus essential.
The projected physical separation between the inner binary is 0.07~au which, at the distance of H36A, translates into a projected angular separation completely unresolvable with the present technology. 
However, the two main components Aa and Ab (average separation of $\sim 4.5$~au) can be resolved (and have been once) using long-baseline optical interferometry. 
By now a unique astrometric observation is available, whereas at least three are required for the orbital inclination to be accurately calculated. Future astrometric data and the computation of the visual orbit will be of value.

H36A has historically called the attention of the researchers due to its likely extreme youth, having been even suggested to be on the zero age main sequence. 
In a parallel paper (Arias et al. 2018, in preparation) we use theoretical models along with the disentangled spectra presented in this work to completely characterize the physical properties of the three stellar components and investigate the evolutionary status of this peculiar massive system.

\bigskip
\section{Acknowledgements}
 We dedicate this work to Nolan R. Walborn, who was a dear colleague, mentor and friend of some of us, and sadly passed away on February 2018. 
The authors thank the reviewer Andrei Tokovinin for his meticulous examination
and many valuable comments which helped to improve this paper.
R.H.B and N.I.M. gratefully acknowledge substantial amounts of observing time granted to "OWN Survey" project by Carnegie Observatories, CNTAC and ESO. 
The authors appreciate the expert support of the LCO staff, which makes observing at Las Campanas a very pleasant experience. 
Also, they gratefully acknowledge the excellent support received from the staff at La Silla. 
Financial support from the Direcci\'on de Investigaci\'on y Desarrollo de la Universidad de La Serena through projects PT14144 and PR16142 is aknowledged by A.R.C. and J.I.A., respectively. 
This work has made use of data from the European Space Agency (ESA) mission {\it Gaia} ({\tt https://www.cosmos.esa.int/gaia}), processed by the 
{\it Gaia} Data Processing and Analysis Consortium (DPAC, {\tt https://www.cosmos.esa.int/web/gaia/dpac/consortium}). Funding for the DPAC has been provided by national institutions, in particular the institutions participating in the {\it Gaia} Multilateral Agreement. 
J.M.A. acknowledges support from the Spanish Government Ministerio de Ciencia, Innovaci\'on y Universidades through grant AYA2016-75\,931-C2-2-P.

%###################################### TABLES
\newpage

%\begin{sidewaystable}
%\begin{landscape}
\begin{table*}
\begin{center}
  \footnotesize
%  \centering
 \caption{Observed radial velocities for the three components of Herschel 36 A used for the inner and outer orbital solutions with \texttt{orbit3.pro} code. In the outer orbit, the O$-$C values for the component Aa1 represent the observed minus calculated RV values; while  for components Ab1 and Ab2, the O$-$C values are the derived barycentric values for orbits of each component minus the calculated values. In the case of the inner orbit, radial velocities marked with $\dag$ represent RVs of Ab1 and Ab2 components corrected by the barycenter motion in the outer orbit.}
  \begin{tabular}{ c   rrrrrr c  rrrrrr }
\hline
\hline
   & \multicolumn{6}{c}{Outer orbit} & & \multicolumn{6}{c}{Inner orbit} \\
   \cline{2-7} \cline{9-14} \\
%   & \multicolumn{6}{c}{-------------------------------------------------------------------------} & \multicolumn{6}{c}{\line} \\
HJD& RV& (O$-$C)& RV& (O$-$C)& RV& (O$-$C)&   & RV& RV$^{\dag}$& (O$-$C)$^{\dag}$&    RV& RV$^{\dag}$& (O$-$C)$^{\dag}$\\
($-$2.400.000)& \Ct& \Ct& BC$_{\text{\Cp}}$& BC$_{\text{\Cp}}$& BC$_{\text{\Cs}}$& BC$_{\text{\Cs}}$& & \Cp& \Cp& \Cp& \Cs& \Cs& \Cs \\
\hline
53873.9166&  $-$56.7&     2.3&     50.9&    1.2&        -&       -&&     126.0&      76.3&    1.0&         -&         -&       -\\
53875.9221&  $-$52.4&     5.2&     50.0&    1.3&        -&       -&&     182.7&     134.0&    0.6&         -&         -&       -\\
53920.6423&  $-$28.2&  $-$4.8&     26.1&    0.6&        -&       -&&     151.2&     125.7&    0.1&         -&         -&       -\\
53936.6823&  $-$11.4&     0.6&     15.4& $-$2.4&        -&       -&&  $-$161.6&  $-$179.4& $-$2.3&         -&         -&       -\\
53937.6292&  $-$11.9&  $-$0.5&     18.0&    0.7&        -&       -&&     126.4&     109.1&    0.2&         -&         -&       -\\
53967.5807&      5.9&  $-$1.0&      3.8& $-$1.1&      7.2&     2.3&&  $-$156.4&  $-$161.3& $-$0.9&     268.7&     263.8&     2.0\\
54210.8726&     37.0&     1.3&   $-$9.7&    5.1&  $-$11.1&     3.7&&  $-$154.7&  $-$139.9&    5.1&     225.6&     240.4&     3.8\\
54210.9020&     36.1&     0.4&  $-$11.9&    2.9&  $-$14.8&  $-$0.0&&  $-$168.8&  $-$154.0&    2.9&     241.3&     256.1&     0.0\\
54211.8558&     34.3&  $-$0.9&  $-$17.4& $-$3.0&  $-$14.3&     0.1&&     159.1&     173.5& $-$3.2&  $-$302.4&  $-$288.0&     0.1\\
54258.8971&   $-$3.7&  $-$1.2&     11.6&    0.2&        -&       -&&  $-$160.3&  $-$171.7&    0.3&         -&         -&       -\\
54258.9087&   $-$3.6&  $-$1.0&      7.4& $-$4.0&        -&       -&&  $-$161.6&  $-$173.0& $-$3.9&         -&         -&       -\\
54599.7856&     51.1&     0.4&  $-$27.9& $-$2.9&  $-$24.5&     0.5&&  $-$103.4&   $-$78.4& $-$2.7&      98.7&     123.7&     0.4\\
54600.7773&     50.8&  $-$0.0&  $-$25.0&    0.1&  $-$24.3&     0.8&&  $-$107.0&   $-$81.9&    0.1&     109.6&     134.7&     0.9\\
54601.7546&     49.6&  $-$1.3&  $-$25.0&    0.2&  $-$27.6&  $-$2.4&&     150.0&     175.2&    0.1&  $-$313.2&  $-$288.0&  $-$2.4\\
54601.8789&     51.4&     0.5&  $-$25.2&    0.0&  $-$24.9&     0.3&&     150.3&     175.5& $-$0.1&  $-$311.3&  $-$286.1&     0.3\\
54625.6767&     54.1&     2.0&  $-$27.8& $-$1.8&  $-$29.1&  $-$3.1&&  $-$206.9&  $-$180.9& $-$1.7&     263.2&     289.2&  $-$3.1\\
54625.7851&     51.4&  $-$0.7&  $-$28.0& $-$2.0&  $-$24.7&     1.3&&  $-$201.1&  $-$175.1& $-$1.8&     257.9&     283.9&     1.3\\
54625.8946&     53.1&     1.0&  $-$26.0&    0.0&  $-$28.7&  $-$2.7&&  $-$159.3&  $-$133.3&    0.2&     189.0&     215.0&  $-$2.7\\
54626.6375&     53.0&     0.9&  $-$22.7&    3.3&  $-$28.7&  $-$2.7&&     123.7&     149.7&    3.2&  $-$267.6&  $-$241.6&  $-$2.6\\
54626.8825&     50.2&  $-$1.9&  $-$26.5& $-$0.5&  $-$24.9&     1.1&&   $-$36.7&   $-$10.7& $-$0.5&    $-$8.2&      17.8&     1.2\\
54627.6221&     51.3&  $-$0.8&  $-$25.4&    0.6&  $-$28.2&  $-$2.2&&   $-$38.1&   $-$12.1&    0.7&    $-$7.4&      18.6&  $-$2.3\\
54670.6413&     44.3&  $-$3.2&  $-$18.4&    4.5&        -&       -&&  $-$129.4&  $-$106.5&    4.6&         -&         -&       -\\
54670.7486&     44.0&  $-$3.5&  $-$17.2&    5.6&        -&       -&&   $-$57.3&   $-$34.5&    5.7&         -&         -&       -\\
54671.6735&     44.8&  $-$2.5&  $-$26.5& $-$3.8&        -&       -&&   $-$97.8&   $-$75.1& $-$3.8&         -&         -&       -\\
54671.7801&     44.2&  $-$3.0&  $-$18.6&    4.1&        -&       -&&  $-$153.3&  $-$130.6&    4.1&         -&         -&       -\\
54954.8244&      8.8&     1.0&      5.4&    1.2&      7.4&     3.2&&     186.2&     182.0&    0.9&  $-$287.6&  $-$291.8&     3.0\\
54955.6602&      8.1&  $-$0.1&      0.6& $-$3.3&      3.6&  $-$0.3&&  $-$176.4&  $-$180.3& $-$3.3&     292.5&     288.6&  $-$0.4\\
56498.6207&     37.1&     1.6&  $-$11.8&    2.9&        -&       -&&  $-$178.2&  $-$163.5&    3.1&         -&         -&       -\\
56498.6445&     37.2&     1.7&  $-$11.4&    3.3&        -&       -&&  $-$183.9&  $-$169.2&    3.5&         -&         -&       -\\
56498.6678&     37.0&     1.5&  $-$16.7& $-$2.0&        -&       -&&  $-$193.6&  $-$178.9& $-$1.8&         -&         -&       -\\
56813.8331&  $-$68.3&     0.2&     53.8& $-$2.3&        -&       -&&     207.0&     150.9& $-$3.1&         -&         -&       -\\
56813.8564&  $-$67.8&     0.7&     54.0& $-$2.1&        -&       -&&     215.6&     159.5& $-$2.9&         -&         -&       -\\
56813.8796&  $-$67.9&     0.6&     54.5& $-$1.6&        -&       -&&     223.1&     167.0& $-$2.3&         -&         -&       -\\
56816.8360&  $-$68.4&  $-$1.2&     59.7&    4.5&        -&       -&&     173.7&     118.5&    3.6&         -&         -&       -\\
56816.8596&  $-$67.9&  $-$0.7&     55.9&    0.8&        -&       -&&     182.9&     127.8& $-$0.2&         -&         -&       -\\
56816.8833&  $-$68.8&  $-$1.6&     59.4&    4.3&        -&       -&&     198.3&     143.2&    3.4&         -&         -&       -\\
57116.7975&     52.6&     2.0&  $-$24.8&    0.1&        -&       -&&  $-$193.2&  $-$168.3&    0.5&         -&         -&       -\\
57570.7970&     51.6&  $-$0.2&  $-$28.0& $-$2.2&        -&       -&&     142.3&     168.1& $-$2.5&         -&         -&       -\\
\hline
\hline
  \end{tabular}
  \label{tvrall}
%\end{sidewaystable}
\end{center}
\end{table*}
%\end{landscape}
\normalsize


\begin{thebibliography}{99}

\bibitem[Allen, 1986]{allen86}
Allen, D.A. 1986, MNRAS, 219P, 35

\bibitem[Andersen \& Clausen, 1989]{andersen89}
Andersen, J., Clausen J.V. 1989, \aap, 213, 183

\bibitem[Antognini \& Thompson, 2016]{antognini16}
Antognini, Joseph M. O., Thompson, Todd A., 2016, \mnras, 456, 4219.

\bibitem[Arias et al., 2006]{arias06}
Arias, J.~I., Barbá, R.~H., {Maíz Apellániz}, J., Morrell, N.~I., Rubio, M., 2006, \mnras, 366, 739.

\bibitem[Arias et al., 2007]{arias07}
Arias, J.~I., Barbá, R.~H., Morrell, N.~I., 2007, \mnras, 374, 1253.

\bibitem[Arias et al., 2010]{arias10}
Arias, J.~I., Barbá, R.~H., Gamen, R.~C., Morrell, N.~I., Maíz Apellániz, J., Alfaro, E.~J., Sota, A., Walborn, N.~R., Moni Bidin, C., 2010, \apjl, 710, L30.

\bibitem[Arias et al., 2016]{arias16}
{Arias, J. I., Walborn, N. R., Simón Díaz, S., Barbá, R. H., Maíz Apellániz, J., Sabín-Sanjulián, C., Gamen, R. C.,  Morrell, 
N. I., Sota, A., Marco, A., Negueruela, I., Leão, J. R. S., Herrero, A., Alfaro, E. J.}, 2016, {\aj}, 152, 31A.

\bibitem[Barbá et al., 2010]{barba10}
Barbá, R.~H., Gamen, R.~C., Arias, J.~I., Morrell, N.~I, Maíz Apellániz, J., Alfaro, E., Walborn, N., Sota, A., 2010, \rmxaa, 38, 30.

\bibitem[Barbá et al., 2017]{barba17}
Barbá, R. H., Gamen, R., Arias, J. I., Morrell, N. I., 2017, IAU Symposium, 329, 89.

\bibitem[Bate et al., 2010]{bate10}
Bate M. R., Lodato G., Pringle J. E., 2010, \mnras, 401, 1505.

\bibitem[Bate, 2014]{bate14}
Bate, Matthew R., 2014, \mnras, 442, 285.

\bibitem[Burkholder et al., 1997]{burkholder97}
Burkholder, V., Massey, P., Morrell, N., 1997, \apj, 490, 328. 

\bibitem[Ferrero et al., 2013]{ferrero13}
{Ferrero, G., Gamen, R., Benvenuto, O., Fern\'andez-Laj\'us, E.} 2013, \mnras, 433, 1300.

\bibitem[Freyhammer et al., 2001]{freyhammer01}
Freyhammer, L.~M., Clausen, J.~V., Arentoft, T., Sterken, C. 2001, \aap, 369, 561.

\bibitem[Gaia Collaboration, 2018]{gaia18a}
{Gaia Collaboration}; Brown, A.~G.~A. et al. 2018, \aap, 616, A1.

\bibitem[Gonz\'alez \& Levato, 2006]{gl06}
Gonz\'alez, J. F., Levato, H., 2006, \aap, 448, 283.

\bibitem[Gonz\'alez et al., 2014]{gonzalez14}
Gonz\'alez, J. F., Veramendi, M.E., Cowley, C.R. 2014, \mnras, 443, 1523

\bibitem[Goto et al., 2006]{goto06}
Goto, M., Stecklum, B., Linz, H., Feldt, M., Henning, T., Pascucci, I., Usuda, T., 2006, \apj, 649, 299

%\bibitem[Hosokawa \& Omukai, 2009]{hosokawa09}
%{Hosokawa, T, Omukai, K}, 2009, {\apj}, 691, 823.

\bibitem[Herschel, 1847]{herschel1847}
Herschel, J. F. W., Sir, "Results of astronomical observations made during the years 1834, 5, 6, 7, 8, at the Cape of Good Hope; being the completion of a telescopic survey of the whole surface of the visible heavens, commenced in 1825", London, Smith, Elder and co., 1847.

\bibitem[Jayasinghe et al., 2018]{jayasinghe18}
Jayasinghe, T., Stanek, K. Z., Kochanek, C. S., et al. 2018, arXiv:1809.07329.

\bibitem[Kochanek et al., 2017]{kochanek17}
Kochanek, C. S., Shappee, B. J., Stanek, K. Z. 2017, PASP, 129, 4502

\bibitem[Krumholz et al., 2009]{krumholz09}
Krumholz, Mark R., Klein, Richard I., McKee, Christopher F., Offner, Stella S. R., Cunningham, Andrew J., 2011, Sci, 323, 754


\bibitem[Henden et al., 2015]{henden15}
Henden, A. A., Levine, S., Terrell, D., Welch, D. L. 2015, American Astronomical Society Meeting Abstracts \#225, 225, id.336.16.


%\bibitem[Le Bouquin et al., 2017]{lebouquin17}
%Le Bouquin, J.-B., Sana, H., Gosset, E., De Becker, M., Duvert, G., Absil, O., Anthonioz, F., Berger, J.-P., Ertel, S., Grellmann, R., Guieu, S., Kervella, P., Rabus, M., Willson, M., 2017, \aap, 601, 34.

\bibitem[Lindegren et al., 2018a]{Lindetal18a}
Lindegren, L. et al., 2018, \aap, 616, A2.

\bibitem[Lindegren et al., 2018b]{Lindetal18b}
Lindegren, L. et al., 2018, {\tt https://www.cosmos.esa.int/documents/29201/1770596/Lindegren\_GaiaDR2\_Astrometry\_extended.pdf/1ebddb25-f010-6437-cb14-0e360e2d9f09}

\bibitem[Luri et al., 2018]{luri18}
Luri, X., et al. 2018, \aap, 616, A9.

\bibitem[Lutz \& Kelker, 1973]{LutzKelk73}
Lutz, T. E. and Kelker, D. H., 1973, \pasp, 85, 573.

\bibitem[Mahy et al., 2018]{mahy18}
Mahy, L., Gosset, E., Manfroid, J. et al. 2018, \aap, 616, A75

\bibitem[Maíz Apellániz, 2001]{Maiz01a}
Ma{\'{\i}}z Apell{\'a}niz, J., 2001, \aj, 121, 2737.

\bibitem[Maíz Apellániz, 2004]{maiz04}
Maíz Apellániz, J. 2004, \pasp, 116, 859.

\bibitem[Maíz Apellániz, 2005]{Maiz05c}
Ma{\'{\i}}z Apell{\'a}niz, J., 2005, in "The Three-Dimensional Universe with Gaia", ESA Special Publication, 576, Turon, C. and O'Flaherty, K.~S. and Perryman, M.~A.~C. (eds.), 179.

\bibitem[Maíz Apellániz et al., 2008]{Maizetal08a}
Ma{\'{\i}}z Apell{\'a}niz, J. and Alfaro, E.~J. and Sota, A., 2008, arXiv:0804.2553.

\bibitem[Maíz Apellániz et al., 2012]{maiz12}
Maíz Apellániz, J., Pellerin, A., Barbá, R. H., et al. 2012, ASP Conf. Ser., 465, 484.

\bibitem[Maíz Apellániz et al., 2015a]{maiz15a}
Maíz Apellániz, J., Úbeda, L., Barbá, R.~H., MacKenty, J.~W., Arias, J.~I., Gómez de Castro, A.~I., 2015, Highlights of Spanish Astrophysics VIII, 604.

\bibitem[Maíz Apellániz et al., 2015b]{maiz15b}
Maíz Apellániz, J., Alfaro, E.~J., Arias, J.~I., Barbá, R.~H., Gamen, R.~C., Herrero, A., Leão, J.~R.~S., Marco, A., Negueruela, I., Simón-Díaz, S., Sota, A., Walborn, N.~R., 2015, Highlights of Spanish Astrophysics VIII, 603.

%http://adsabs.harvard.edu/abs/2015hsa8.conf..603M


\bibitem[Maíz Apellániz et al., 2016]{maiz16}
Maíz Apellániz, J., Sota, A., Arias, J. I., Barbá, R. H., Walborn, N. R., Simón-Díaz, S., Negueruela, I., Marco, A., Le\~ao, J. R. S., Herrero, A., Gamen, R. C., Alfaro, E. J., 2016, \apjs, 224, 4.

\bibitem[Mahy et al., 2012]{mahy12}
Mahy, L., Gosset, E., Sana, H., Damerdji, Y., De Becker, M., Rauw, G., Nitschelm, C., 2012, \aap, 540, 97.

\bibitem[Mardling \& Aarseth, 2001]{mardling01}  
Mardling R. A., Aarseth S. J., 2001, \mnras, 321, 398

\bibitem[Martins et al., 2005]{martins05}
Martins, F., Schaerer, D., Hillier, D.~J., 2005, \aap, 436, 1049.

\bibitem[Martins et al., 2017]{martins17}
Martins, F., Mahy, L., Hevré, A. 2017, \aap, 607, A82.

\bibitem[Moe \& Kratter, 2018]{moe18}
Moe, Maxwell, Kratter, Kaitlin M., 2018, \apj, 854, 44.

\bibitem[Moeckel \& Bally, 2007]{moeckel07}
Moeckel, Nickolas, Bally, John, 2007, \apj, 656, 275.

\bibitem[Niemela et al., 2006]{niemela06}
Niemela, V.~S., Morrell, N.~I., Fernández-Lajús, R., Barbá, R.~H., Albacete Colombo, J.~F., Orellana, M., 2006, \mnras, 367, 1450.

\bibitem[Rauw et al., 2001]{rauw01}
Rauw, G., Sana, H., Antokhin, I.~I., Morrell, N.~I., Niemela, V.~S., Albacete COlombo, J.~F., Gosset, E., Vreux, J.-M., 2001, \mnras, 326, 1149.

%\bibitem[Reed, 2003]{reed03}
%Reed, B.C. 2003, AJ, 125, 2531

\bibitem[Sabín-Sanjulián et al., 2014]{sabin14}
Sabín-Sanjulián, C., Simón-Díaz, S., Herrero, A., Walborn, N. R., Puls, J., Maíz Apellániz, J., Evans, C. J., Brott, I., de Koter, A., Garcia, M., Markova, N., Najarro, F., Ramírez-Agudelo, O. H., Sana, H., Taylor, W. D., Vink, J. S., 2014, \aap, 564, A39.

\bibitem[Samus et al., 2017]{samus17}
Samus, N. N., Kazarovets, E. V., Durlevich, O. V., Kireeva, N. N., Pastukhova, E. N. 2017, ARep, 61, 80

\bibitem[Sana et al., 2013]{sana13}
Sana, H., Le Bouquin, J.-B., Mahy, L., Absil, O., De Becker, M., Gosset, E., 2013, \aap, 553, 131.

\bibitem[Sana, 2017]{sana17}
Sana, H., 2017, IAU Symposium, 329, 110.

\bibitem[Sánchez-Bermúdez et al., 2014]{sanchez14}
Sánchez-Bermúdez, J., Alberdi, A., Sch\"odel, R., Hummel, C.~A., Arias, J.~I., Barbá, R.~H., Maíz Apellániz, J., Pott, J. U., 2014, \aap, 572, L1.

\bibitem[Schaefer et al., 2016]{schaefer16}
Schaefer, G. H., Hummel, C. A., Gies, D. R., Zavala, R. T., Monnier, J. D., Walter, F. M., Turner, N. H., Baron, F., ten Brummelaar, T., Che, X., Farrington, C. D., Kraus, S., Sturmann, J., Sturmann, L., 2016, AJ, 152, 213.

%\bibitem[Simón-Díaz et al., 2011]{sd11}
%Simón-Díaz, S., Castro, N., Herrero, A., Puls, J., Garcia, M., Sabín-Sanjulián, C., 2011, {JPhCS}, 328, 2021.

\bibitem[Schmidt, 1868]{schmidt1868}
Schmidt, J. F. J. 1868, AN, 70, 245.

\bibitem[Shappee et al., 2014]{shappee14} 
Shappee, B. J. et al. 2014, \apj, 788, 48.

\bibitem[Simón-Díaz et al., 2011]{sig_ori11}
Simón-Díaz, S., Caballero, J. A., Lorenzo, J., 2011, \apj, 742, 55.

\bibitem[Simón-Díaz et al., 2015]{sig_ori15}
Simón-Díaz, S., Caballero, J. A., Lorenzo, J., Maíz Apellániz, J., Schneider, F. R. N., Negueruela, I., Barbá, R. H., Dorda, R., Marco, A., Montes, D., Pellerin, A., Sanchez-Bermudez, J., Sódor, Á.; Sota, A., 2015, \apj, 799, 169.

\bibitem[Sota et al., 2011]{sota11}
Sota, A., Maíz Apellániz, J. Walborn N. R., Alfaro, E. J., Barbá, R. H., Morrell, N. I., Gamen, R. C., Arias, J. I., 2011, \apjs, 193, 24.

\bibitem[Sota et al., 2014]{sota14}
Sota, A., Maíz Apellániz, J., Morrell, N.~I., Barbá, R.~H., Walborn, N.~R., Gamen, R.~C., Arias, J.~I., Alfaro, E.~J., 2014, \apjs, 211, 10.

\bibitem[Stecklum et al., 1995]{stecklum95}
Stecklum, B., Henning, T., Eckart, A., Howell, R.~R., Hoare, M.~G., 1995, \apjl, 445, 153.

\bibitem[Stecklum et al., 1998]{stecklum98}
Stecklum, B., Henning, T., Feldt, M., Hayward, T. L., Hoare, M. G., Hofner, P., Richter, S., 1998, \aj, 115, 767.

\bibitem[Sung et al., 2000]{sung00}
Sung, H., Chun, M.-Y. , Bessell, M.S. 2000, AJ, 120, 333

\bibitem[Tokovinin, 1992]{tokovinin92}
Tokovinin, A., 1992, Complementary Approaches to Double and Multiple Star Research, ASP Conference Series, Vol. 32, IAU Colloquium 135.

\bibitem[Tokovinin, 2004]{tokovinin04} 
Tokovinin, A., 2004, RMxAC, 21, 7.

\bibitem[Tokovinin \& Latham, 2017]{tokovinin17}
Tokovinin, A., Latham, D.W., 2017, \apj, 838, 54.

\bibitem[Tokovinin, 2017]{tokovinin17st}
Tokovinin, A., 2017, \apj, 844, 103.

\bibitem[Tokovinin, 2018]{tokovinin18st}
Tokovinin, A., 2018, \apjs, 235, 6.

\bibitem[Veramendi, 2012]{veramendi12}
Veramendi, M.~E., 2012, Phd. Thesis, Universidad Nacional de Córdoba, Argentina.

\bibitem[Walborn \& Fitzpatrick, 1990]{walborn90}
Walborn, N.~R., and Fitzpatrick, E.~L., 1990, \pasp, 102, 379.

\bibitem[Walborn, 2007]{walborn07}
Walborn, N.~R., 2007, arXiv:astro-ph/0701573v2. 

\bibitem[Wheelwright et al., 2011]{wheelwright11}
Wheelwright, H. E., Vink, J. S., Oudmaijer, R. D., Drew, J. E., 2011, \aap, 532, 28.

\bibitem[Wood \& Churchwell, 1989]{wood89}
Wood, D.~O.~S., Churchwell, E., 1989, \apjs, 69, 831.

%\bibitem[Woodward et al., 1985]{woodward85}
%Woodward, C. E.,Pipher, J. L., Holfer, H. L., Forrest, W. J., 1985, BAAS, 17, 837.

\bibitem[Woodward et al., 1986]{woodward86}
Woodward, C.~E., Pipher, J.~L., Helfer, H.~L., Sharpless, S., Moneti, A., Kozikowski, D., Oliveri, M., Willner, S.~P., Lacasse, M.~G., Herter, T., 1986, \aj, 91, 870.

\bibitem[Woodward et al., 1990]{woodward90}
Woodward, C.~E., Pipher, J.~L., Helfer, H.~L., Forrest, W.~J., 1990, \apj, 365, 252.


\end{thebibliography}
\end{document}